\documentclass[journal]{IEEEtran}

\usepackage{setspace,amsmath,latexsym,cite,amssymb,amsfonts,enumerate,graphicx,epsfig,bm,color}

\ifCLASSOPTIONcompsoc
\usepackage[caption=false,font=normalsize,labelfont=sf,textfont=sf]{subfig}
\else
\usepackage[caption=false,font=footnotesize]{subfig}
\fi

\newtheorem{lemma}{Lemma}
\newtheorem{proposition}{Proposition}
\newtheorem{remark}{Remark}

\title{Throughput Maximization for Decode-and-Forward Relay Channels with Non-Ideal Circuit Power}

\author{Hengjing~Liang,
        Chuan~Huang,~\IEEEmembership{Member,~IEEE,}
        Zhi~Chen,~\IEEEmembership{Member,~IEEE,}
        and~Shaoqian~Li,~\IEEEmembership{Fellow,~IEEE}

\thanks{This paper was presented in part at IEEE Wireless Communications and Networking Conference 2017, San Francisco, CA, USA, March 2017.}
\thanks{This work was supported in part by the High-Tech Research and Development (863) Program of China under Grant 2015AA01A707 and the National Natural Science Foundation of China under Grant 61501093.}
\thanks{The authors are with the National Key Laboratory of Science and Technology on Communications, University of Electronic Science and Technology of China, Chengdu 610054, China (e-mail: lianghj@hotmail.com; \{huangch, chenzhi, lsq\}@uestc.edu.cn).
Corresponding author: C. Huang.}}

\begin{document}

\maketitle

\begin{abstract}
This paper studies the throughput maximization problem for a three-node relay channel with non-ideal circuit power.
In particular, the relay operates in a half-duplex manner, and the decode-and-forward (DF) relaying scheme is adopted.
Considering the extra power consumption by the circuits, the optimal power allocation to maximize the throughput of the considered system over an infinite time horizon is investigated.
First, two special scenarios, i.e., the direct link transmission (only use the direct link to transmit) and the relay assisted transmission (the source and the relay transmit with equal probability), are studied, and the corresponding optimal power allocations are obtained.
By transforming two non-convex problems into quasiconcave ones, the closed-form solutions show that the source and the relay transmit with certain probability, which is determined by the average power budgets, circuit power consumptions, and channel gains.
Next, based on the above results, the optimal power allocation for both the cases with and without direct link is derived, which is shown to be a mixed transmission scheme between the direct link transmission and the relay assisted transmission.
\end{abstract}


\section{Introduction}
\label{sec_intro}
Green communication has drawn great attention during the past years.
It is reported that more than 1 million gallons of diesel are consumed by Vodafone, for example, to power their cellular networks \cite{CHan11}, and the consumption will still go up in the future.
The growing cost of fossil fuel energy calls for both environmental and economical demands and motivations for the design of green communications \cite{YLi11}.

Circuit energy consumption amounts for a significant part of the total energy consumption \cite{SCui04,SCui05}.
In order to reduce circuit energy consumption for a fixed amount of data transmission, increasing throughput and reducing transmission time are the key targets.
Thus, green communication associated with non-ideal circuit power needs to be designed both energy and spectrum efficiently.
A generic energy efficiency (EE) maximization problem considering circuit power consumption was summarized in \cite{CIsheden12}.
In \cite{GMiao10}, a link adaptation scheme that balances circuit power consumption and transmission power was proposed in frequency-selective channels.
EE maximization problems with circuit energy consumption were also considered in orthogonal frequency division multiple access (OFDMA) \cite{CXiong11} and wireless sensor networks \cite{SKJay11}.
A throughput optimal policy considering circuit power was proposed for point-to-point channels with energy harvesting transmitter \cite{JXu14}.

Relaying has been considered as a promising technique to mitigate fading and extend coverage in wireless networks, which was introduced in \cite{vandermeulen1971} and comprehensively studied in \cite{TMCover1979}.
Decode-and-forward (DF) relaying was studied in \cite{JNLaneman04,RPabst04,ANosratinia04,GKramer05}.
The capacity of a classical three-node relay channel, consisting of a source, a destination, and a single half-duplex DF relay, was investigated in \cite{AHost05}, and the capacity analysis is extended to a parallel fading relay channel in \cite{YLiang07}.
Resource allocation problems maximizing spectral efficiency (SE) for relay networks under different scenarios have been investigated in \cite{TCYNg07,BRankov07,JTang07,OSimeone07,KJitvanichphaibool09,CHuang13}.
Green communication problems in relay networks were discussed in \cite{YYao05,rmadan08,ZZhou08,GBrante11,CBae09,XZhou15,DWang16,szhou11}.
In \cite{YYao05}, minimum energy required to transmit one information bit was studied in amplify-and-forward (AF) and DF.
In \cite{rmadan08} and \cite{ZZhou08}, energy minimization problems considering channel state information acquiring energy and signaling overhead were investigated in single relay selection scheme, respectively.
In \cite{GBrante11} and \cite{CBae09}, non-ideal circuit power consumption, i.e., non-zero circuit power consumption during transmission, was considered for total energy minimization problems in multihop relay channels.
In \cite{XZhou15}, sum rate maximization problem with non-ideal circuit power was studied under holistic power constraints for the multiple-input multiple-output two-way AF relay channels.
In \cite{DWang16}, circuit power consumption was considered for the secure EE maximization of AF relay channels.
In \cite{szhou11}, sleep mode was further introduced to save energy in a one-dimension cellular network, where the relay placement and the relay sleep probability were jointly optimized.
In \cite{HLwcnc}, throughput maximization problems with non-ideal circuit power consumption were studied in a three-node relay channel with direct link.

In this paper, throughput maximization for a three-node half-duplex Gaussian relay channels considering non-ideal circuit power is studied over an infinite time horizon.
The transceiver circuitry consumes a constant amount of power in the active mode and negligible power in the sleep mode.
Under this setup, the optimal power allocations for the throughput maximization of the relay channel with and without direct link are both investigated.
The main contributions of this paper are summarized as follows.
\begin{itemize}
\item First, the throughput maximization problems for two special scenarios are investigated.
    For the direct link transmission (DLT), where only direct link is used for transmission, the optimal power allocation shows that the source transmits either at a certain portion of time slots or constantly according to different average power budgets, circuit power consumptions, and channel power gains.
    Then, the average throughput for DLT is obtained.
    For the relay assisted transmission with direct link (RAT-DL), where the source and the relay transmit with equal probability, the optimal power allocation has a similar transmission structure as DLT.
    By solving a max-min problem, the average throughput for RAT-DL is obtained.
\item Then, with the two special power allocation cases (DLT and RAT-DL) and the characteristics of their average throughputs, the optimal power allocation for the throughput maximization of the relay channel with the direct link, where the source and the relay are not constrained to transmit with equal probability, is studied.
    The optimal solutions obtained by graphic method are shown to be either a single type of transmission (DLT or RAT-DL) or a time sharing of both transmissions.
    Whether to choose RAT-DL depends on the average power budget and the maximum EE of DLT and RAT-DL.
\item Furthermore, the optimal power allocation for the relay assisted transmission without direct link (RAT-WDL), where the source and the relay transmit with equal probability and the direct link is inactive, is analyzed.
    Asymptotic analysis is given for DLT, RAT-DL together with RAT-WDL at the low signal-to-noise ratio (SNR) and high SNR regimes afterwards.
    At last, simulation results show that the optimal power allocation scheme outperforms other conventional schemes.
\end{itemize}

The rest of this paper is organized as follows.
Section \ref{sec_sys_mdl} introduces the system model and the main assumptions of this paper.
Section \ref{sec_2spe} studies two special scenarios.
Section \ref{sec_opt} investigates the optimal power allocation scheme for the case with direct link and asymptotic performances.
Section \ref{sec_relaywod} analyzes RAT-WDL and Section \ref{sec_sim} evaluates the throughput performances by simulations.
Finally, Section \ref{sec_conclu} concludes the paper.

\emph{Notation}: $\mathcal{C}\left( x \right)={{\log }_{2}}\left( 1+x \right)$ denotes the capacity of the additive white Gaussian noise (AWGN) channel, where $x$ is the SNR of the channel.

\section{System Model}
\label{sec_sys_mdl}
\begin{figure}[!t]
\centering
\includegraphics[width=3in]{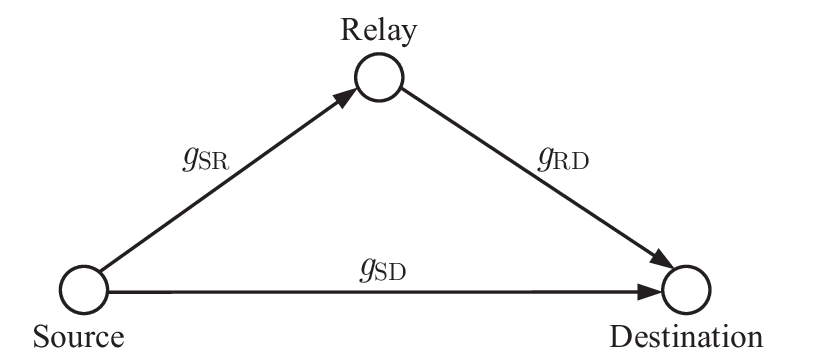}
\caption{A three-node relay channel.}
\label{fig_sys}
\end{figure}

This paper considers a three-node relay channel as shown in Fig. \ref{fig_sys}, which consists of a source, a destination, and a half-duplex relay.
The source sends information to the destination with the help of the relay.
Slotted transmission scheme is adopted, and each time slot is with duration $T$.

\subsection{Signal Model}
\label{sec_sys_mdl_signal_model}
In this subsection, channel input and output relationship of the considered relay channel is introduced.
Denote the channel coefficients of the source-destination, source-relay, and relay-destination links as $g_\text{SD}$, $g_\text{SR}$, and $g_\text{RD}$, respectively, and then the channel power gains of the three links are given by
\begin{equation}
{{h}_{\text{SD}}}={{\left| {{g}_{\text{SD}}} \right|}^{2}},\quad {{h}_{\text{SR}}}={{\left| {{g}_{\text{SR}}} \right|}^{2}},\quad {{h}_{\text{RD}}}={{\left| {{g}_{\text{RD}}} \right|}^{2}},
\end{equation}
which are all constants across the time slots.

If the relay is not selected to help the source transmission, the received signal at the destination in time slot $i$ is given as
\begin{equation}
y_\text{D}\left( i \right)={{g}_{\text{SD}}}x\left( i \right)+n_\text{D}\left( i \right),
\end{equation}
where $x\left( i \right)$ is the source transmitted signal with power ${{P}_{\text{S}}}\left( i \right)$, and $n_\text{D}\left( i \right)$ is the independent and identically distributed (i.i.d.) circularly symmetric complex Gaussian (CSCG) noise with zero mean and unit variance.

When the DF relaying scheme is adopted to help the source transmission, it operates in a half-duplex manner (one time slot is then divided into two phases), and the information encoding and decoding processes are described as follows:
\begin{enumerate}
\item In the first phase of time slot $i$, the source broadcasts $x\left( i \right)$ to both the relay and the destination with power ${{P}_{\text{S}}}\left( i \right)$;
\item Then, the received signal at the relay during the first phase of time slot $i$ is given as
\begin{equation}
y_\text{R}\left( i \right)={{g}_{\text{SR}}}x\left( i \right)+n_\text{R}\left( i \right),
\end{equation}
where $n_\text{R}\left( i \right)$ is i.i.d. CSCG noise with zero mean and unit variance.
Next, the relay decodes the source message, re-encodes it into a new signal $\tilde{x}\left( i \right)$, and forwards $\tilde{x}\left( i \right)$ to the destination with power ${{P}_{\text{R}}}\left( i \right)$.
\item Finally, the destination receives the signals over the whole time slot, and the received signals $y_\text{D}^1\left( i \right)$ and $y_\text{D}^2\left( i \right)$ in the two phases are given as
\begin{eqnarray}
y_\text{D}^1\left( i \right)={{g}_{\text{SD}}}x\left( i \right)+n_\text{D}^1\left( i \right),\\
y_\text{D}^2\left( i \right)={{g}_{\text{RD}}}\tilde{x}\left( i \right)+n_\text{D}^2\left( i \right),
\end{eqnarray}
respectively, where $n_\text{D}^1\left( i \right)$ and $n_\text{D}^2\left( i \right)$ are i.i.d. CSCG noise with zero mean and unit variance.
\end{enumerate}

For the purpose of exposition, consider the case that the two phases in one time slot are with equal length.
Thus, the transmission rate for the DF relaying scheme at time slot $i$ is given as \cite{AHost05}
\begin{align}
\nonumber
{{R}}_\text{B}\left( i \right)=&\frac {1}{2}\min \left\{  \mathcal{C}\left( {{P}_{\text{S}}}\left( i \right){{h}_{\text{SR}}} \right),\right.\\
\label{rate_b}
&\left.\mathcal{C}\left({{P}_{\text{S}}}\left( i \right){{h}_{\text{SD}}} \right)+\mathcal{C}\left({{P}_{\text{R}}}\left( i \right){{h}_{\text{RD}}} \right) \right\}.
\end{align}
It is well known that the DF relaying scheme can work only when $h_\text{SR} \ge 2 h_\text{SD}$ \cite{AHost05}; otherwise, DLT without the help of relay achieves a larger rate.

\subsection{Power Consumption Model}
In this subsection, power consumption model considering the non-ideal circuit power is discussed.
The transceiver circuitry works in two modes: when a signal is transmitting, all circuits work in the \emph{active mode}; and when there is no signal to transmit, they work in the \emph{sleep mode}.
\begin{enumerate}
\item \textbf{Active mode}: the consumed power is mainly comprised of the transmission power and the circuit power.
The transmission power is determined by the power allocation ${{P}_{\text{S}}}\left( i \right)$ and ${{P}_{\text{R}}}\left( i \right)$.
The circuit power consists of the following two parts: the transmitting circuit power ${{P}_{\text{ct}}}$ comes from the power consumed by the mixer, frequency synthesizer, active filter, and digital-to-analog converter \cite{SCui05}; and the receiving circuit power ${{P}_{\text{cr}}}$ is composed of the power consumption of the mixer, frequency synthesizer, low noise amplifier, intermediate frequency amplifier, active filter, and analog-to-digital converter \cite{SCui05}.
Constant circuit power model is considered in this paper, i.e., ${{P}_{\text{ct}}}$ and ${{P}_{\text{cr}}}$ are constants \cite{SCui05}.
In the sequel, superscripts ``S'', ``R'', and ``D'' are added to ${{P}_{\text{ct}}}$ and ${{P}_{\text{cr}}}$ to distinguish the power consumed at the source, relay, and destination, respectively.
\item \textbf{Sleep mode}: it has been shown that the power consumption ${{P}_{\text{sp}}}$ in the sleep mode is dominated by the leaking current of the switching transistors and is usually much smaller than that in the active mode \cite{SCui05}.
    Therefore, the power consumption in the sleep mode is set as ${{P}_{\text{sp}}}=0$. It is worth pointing out that the results of this paper can be readily extended to the case of ${{P}_{\text{sp}}}\ne 0$ by deducting ${{P}_{\text{sp}}}$ from the average power budget and the power consumption in the active mode.
\end{enumerate}

In general, the circuit power consumed in the active mode is larger than that in the sleep mode, i.e.,
\begin{equation}
{{P}_{\text{cr}}}>{{P}_{\text{ct}}}>{{P}_{\text{sp}}}.
\end{equation}
Thus, smartly operating between the two modes can potentially save a significant amount of energy.

Based on the power model discussed above, the power consumptions for both DLT and RAT-DL are computed as follows.
\begin{enumerate}
\item DLT: Denote $\alpha_\text{A}$ as the total circuit power consumption in the active mode for DLT, and it is the sum of the transmitting circuit power at the source and the receiving circuit power at the destination, i.e.,
\begin{equation}
\alpha_\text{A}={{P}_{\text{ct}}^\text{S}}+{{P}_{\text{cr}}^\text{D}}.
\end{equation}
With the defined $\alpha_\text{A}$ and ${{P}_{\text{sp}}}=0$, the total power consumption at time slot $i$ for DLT is thus given as
\begin{equation}
P^\text{A}_\text{total}\left( i \right)= \begin{cases}
   0 & {{P}_{\text{S}}}\left( i \right)=0  \\
   {{P}_{\text{S}}}\left( i \right)+\alpha_\text{A} & {{P}_{\text{S}}}\left( i \right)>0.  \\
\end{cases}
\end{equation}

Then, the average power constraint for DLT is defined over $N$ time slots, as $N$ goes to infinity, i.e.,
\begin{equation}
\label{d_total_power_cons}
\underset{N\to \infty }{\mathop{\lim }}\,\frac{1}{N}\sum\limits_{i=1}^{N}P^\text{A}_\text{total}\left(i\right)\le P_\text{A},
\end{equation}
where $P_\text{A}\ge 0$ is the power budget.

\item RAT-DL: Denote $\alpha_\text{B}$ as the total circuit power consumption in the active mode for the transmission with the help of a relay, and it is the sum of the transmitting circuit power at the source and the relay, the receiving circuit power at the relay and the destination, i.e.,
\begin{equation}
\label{alpha_r}
\alpha_\text{B}=\frac{1}{2}\left( {{P}_{\text{ct}}^\text{S}}+{{P}_{\text{cr}}^\text{R}}+{{P}_{\text{cr}}^\text{D}} \right)+\frac{1}{2}\left( {{P}_{\text{ct}}^\text{R}}+{{P}_{\text{cr}}^\text{D}} \right),
\end{equation}
where the $\frac{1}{2}$ penalty is due to the half-duplex constraint for the considered relaying scheme.
With the defined $\alpha_\text{B}$ and ${{P}_{\text{sp}}}=0$, the total power consumption at time slot $i$ for RAT-DL is given as
\begin{equation}
\label{P_total^R}
P^\text{B}_\text{total}\left( i \right)=\begin{cases}
   0 & {{P}_{\text{S}}}\left( i \right)=0,{{P}_{\text{R}}}\left( i \right)=0  \\
   \frac{{{P}_{\text{S}}}\left( i \right)+{{P}_{\text{R}}}\left( i \right)}{2}+\alpha_\text{B} & {{P}_{\text{S}}}\left( i \right)>0,{{P}_{\text{R}}}\left( i \right)>0,  \\
\end{cases}
\end{equation}
where the $\frac{1}{2}$ penalty is also due to the half-duplex constraint for the considered relaying scheme.
Then, the average power constraint for RAT-DL is defined over $N$ time slots, as $N$ goes to infinity, i.e.,
\begin{equation}
\label{r_total_power_cons}
\underset{N\to \infty }{\mathop{\lim }}\,\frac{1}{N}\sum\limits_{i=1}^{N}P^\text{B}_\text{total}\left(i\right)\le P_\text{B},
\end{equation}
where $P_\text{B}\ge 0$ is the power budget.
\end{enumerate}

\section{A Closer Look at Two Special Cases}
\label{sec_2spe}
In this section, the throughput maximization problems for two special scenarios are firstly studied, DLT and RAT-DL, and the corresponding optimal power allocations are obtained for these throughput maximization problems under the two scenarios.

\subsection{Direct Link Transmission}
\label{subsec_dlt}
In this scenario, the source directly transmits to the destination in one whole time slot and the relay is always inactive.
Thus the transmission rate for DLT at time slot $i$ is given as
\begin{equation}
{{R}}\left( i \right)=\mathcal{C}\left( {{P}_{\text{S}}}\left( i \right){{h}_{\text{SD}}} \right).
\end{equation}
The goal is to determine $\left\{ P_\text{S}\left( i \right) \right\}$ such that the long term average throughput subject to the average power constraint defined in (\ref{d_total_power_cons}) is maximized over $N$ time slots as $N\to \infty$, i.e., solve the following optimization problem
\begin{align}
\label{d_link_origin_prb}
\nonumber
\mathcal{C}_\text{A}\left( {{P}_{\text{A}}} \right)=\underset{\left\{ P_\text{S}\left( i \right) \right\}}{\mathop{\max }}\;&\underset{N\to \infty }{\mathop{\lim }}\,\frac{1}{N}\sum\limits_{i=1}^{N}\mathcal{C}\left( {{P}_{\text{S}}}\left( i \right){{h}_{\text{SD}}} \right)\\
\text{s.t. } &(\ref{d_total_power_cons}),\ {{P}_{\text{S}}}\left( i \right)\ge 0.
\end{align}

Similar problem has been studied in \cite{JXu14}.
As the objective function of problem (\ref{d_link_origin_prb}) is nonnegative and concave, its solution is of the same structure as that in \cite{JXu14}, which is summarized as the following lemma.
\begin{lemma}
\label{lem1_direct_link_opt_sol}
The optimal power allocation for problem (\ref{d_link_origin_prb}) is given as: Transmit with power value $P_{\text{S}}^{*}=\max \left( {{P}_{\text{ee1}}},{{P}_{\text{A}}}-\alpha_\text{A} \right)$ over $p^*$ portion of time slots and keep silent for the rest of the slots, where ${{P}_{\text{ee1}}}\triangleq\mathop{\arg}\underset{{{P}_{\text{S}}}>0}{\mathop{\max }}\, \frac{\mathcal{C}\left( {{P}_{\text{S}}}{{h}_{\text{SD}}} \right)}{{{P}_{\text{S}}}+\alpha_\text{A}}$ and ${{p}^{*}}=\frac{{{P}_{\text{A}}}}{P_{\text{S}}^{*}+\alpha_\text{A}}$.
\end{lemma}

\begin{remark}
From Lemma \ref{lem1_direct_link_opt_sol}, it is observed that when the average power budget $P_\text{A}$ is relatively small, i.e., $P_\text{A}\le P_\text{ee1}+\alpha_\text{A}$, the optimal transmission strategy is with an ``on-off'' structure.
It is due to that the scarce average power budget cannot support the constant transmission with non-zero power consumption for the circuits.
Under this circumstance, transmission with power $P_\text{ee1}$ over $\frac{{{P}_{\text{A}}}}{P_{\text{ee1}}+\alpha_\text{A}}$ portion of time slots achieves the maximum transmission throughput.
When the average power budget is large enough, i.e., $P_\text{A}> P_\text{ee1}+\alpha_\text{A}$, the optimal transmission strategy follows a constant transmission with the power value ${{P}_{\text{A}}}-\alpha_\text{A}$.
\end{remark}

With the obtained optimal transmission power $P_{\text{S}}^{*}$ and probability $p^*$ in Lemma \ref{lem1_direct_link_opt_sol}, the relationship between the average throughput $\mathcal{C}_\text{A}\left(P_\text{A}\right)$ defined in (\ref{d_link_origin_prb}) and the average power budget $P_\text{A}$ is summarized in the following proposition.
\begin{proposition}
\label{prp1_CDPA_char}
The average throughput $\mathcal{C}_\text{A}\left(P_\text{A}\right)$ defined in (\ref{d_link_origin_prb}) is given as
\begin{equation}
\label{CDPA}
\mathcal{C}_\text{A}\left( {{P}_{\text{A}}} \right)=\begin{cases}
   \frac{\mathcal{C}\left( {{P}_{\text{ee1}}}{{h}_{\text{SD}}} \right)}{{{P}_{\text{ee1}}}+\alpha_\text{A}}\cdot {{P}_{\text{A}}} & 0\le {{P}_{\text{A}}}\le {{P}_{\text{ee1}}}+\alpha_\text{A}  \\
   \mathcal{C}\left( \left( {{P}_{\text{A}}}-\alpha_\text{A} \right){{h}_{\text{SD}}} \right) & {{P}_{\text{A}}}>{{P}_{\text{ee1}}}+\alpha_\text{A},  \\
\end{cases}
\end{equation}
which is continuous, differentiable, and concave over $P_\text{A}\ge 0$.
\end{proposition}
\begin{IEEEproof}
Please see Appendix \ref{prf_prp1_CDPA_char}.
\end{IEEEproof}

\begin{remark}
From Proposition \ref{prp1_CDPA_char}, it is observed that the average throughput $\mathcal{C}_\text{A}\left(P_\text{A}\right)$ is a linear function of the average power budget $P_\text{A}$ when $P_\text{A}$ is small, i.e., $P_\text{A}\le P_\text{ee1}+\alpha_\text{A}$.
Since ${{P}_{\text{ee1}}}\triangleq\mathop{\arg}\underset{{{P}_{\text{S}}}>0}{\mathop{\max }}\,\frac{\mathcal{C}\left( {{P}_{\text{S}}}{{h}_{\text{SD}}} \right)}{{{P}_{\text{S}}}+\alpha_\text{A}}$, it suggests that the transmission scheme given in Lemma \ref{lem1_direct_link_opt_sol} achieves the maximum EE for the case of DLT \cite{JXu14}.
When the average power budget is large enough, i.e., $P_\text{A}> P_\text{ee1}+\alpha_\text{A}$, the source transmits constantly to achieve the maximum SE.
\end{remark}

\subsection{Relay Assisted Transmission with Direct Link}
\label{subsec_rat}
The optimal power allocation for RAT-DL is studied in this subsection.
In this scenario, the relay works following the DF relaying scheme described above: The source and the relay transmit in each half of the time slot, i.e., the transmission probabilities of the source and the relay are the same.
\subsubsection{Problem Formulation}
The goal is to determine $\left\{ P_\text{S}\left( i \right) \right\}$ and $\left\{{{P}_{\text{R}}}\left( i \right)\right\}$ such that the long term average throughput subject to the average power constraint defined in (\ref{r_total_power_cons}) is maximized over $N$ time slots as $N\to \infty$, i.e., solve the following optimization problem
\begin{align}
\label{r_link_origin_prb}
\nonumber
\mathcal{C}_\text{B}\left( {{P}_{\text{B}}} \right)=\underset{\left\{{{P}_{\text{S}}}\left( i \right)\right\},\left\{{{P}_{\text{R}}}\left( i \right)\right\}}{\mathop{\max }}\;&\underset{N\to \infty }{\mathop{\lim }}\,\frac{1}{N}\sum\limits_{i=1}^{N}R_\text{B}\left(i\right)\\
\text{s.t. }\;\;\;\;\;\;\;&(\ref{r_total_power_cons}),\ {{P}_{\text{S}}}\left( i \right)\ge 0,\ {{P}_{\text{R}}}\left( i \right)\ge 0,
\end{align}
where $R_\text{B}\left(i\right)$ is given in (\ref{rate_b}).

Since the objective function of problem (\ref{r_link_origin_prb}) is nonnegative and concave \cite{SBoyd04}, it can be checked \cite{JXu14} that the optimal power allocation of problem (\ref{r_link_origin_prb}) is given as: Transmit with power ${{P}_{\text{S}}}\left( i \right)={{P}_{\text{S}}}>0$ and ${{P}_{\text{R}}}\left( i \right)={{P}_{\text{R}}}>0$ over $p$ portion of time slots and keep silent for the rest of the slots, where $P_\text{S}$ and $P_\text{R}$ are constants.
As a result, problem (\ref{r_link_origin_prb}) can be reformulated as
\begin{align}
\nonumber
\mathcal{C}_\text{B}\left( {{P}_{\text{B}}} \right)=\underset{\left\{{{P}_{\text{S}}},{{P}_{\text{R}}},p\right\}}{\mathop{\max }}\ &\frac{p}{2}\min \left\{ \mathcal{C}\left( {{P}_{\text{S}}}{{h}_{\text{SR}}} \right),\right.\\
\label{r_link_trans_prb}
&\left.\mathcal{C}\left( {{P}_{\text{S}}}{{h}_{\text{SD}}} \right)+\mathcal{C}\left( {{P}_{\text{R}}}{{h}_{\text{RD}}} \right) \right\}\\
\label{r_link_trans_prb_con1}
\text{s.t. }\;\;\; &\left( \frac{1}{2}{{P}_{\text{S}}}+\frac{1}{2}{{P}_{\text{R}}}+\alpha_\text{B} \right)\cdot p\le {{P}_{\text{B}}},\\
\label{r_link_trans_prb_con2}
&0\le p\le 1,\ {{P}_{\text{S}}}\ge0,\ {{P}_{\text{R}}}\ge0,
\end{align}
where (\ref{r_link_trans_prb_con1}) is obtained from (\ref{r_total_power_cons}).

It is easy to check that to achieve the optimal value of problem (\ref{r_link_trans_prb})--(\ref{r_link_trans_prb_con2}), constraint (\ref{r_link_trans_prb_con1}) must be satisfied with equality.
Thus it follows that the optimal transmission probability $p^*=\frac{2{{P}_{\text{B}}}}{{{P}_{\text{S}}}+{{P}_{\text{R}}}+2\alpha_\text{B}}$. Hence, problem (\ref{r_link_trans_prb})--(\ref{r_link_trans_prb_con2}) can be simplified as
\begin{align}
\nonumber
&\mathcal{C}_\text{B}\left( {{P}_{\text{B}}} \right)=\\
\label{r_link_trans_prb_quasi}
\underset{\left\{{{P}_{\text{S}}},{{P}_{\text{R}}}\right\}}{\mathop{\max }}&\frac{{{P}_{\text{B}}}\min \left\{ \mathcal{C}\left( {{P}_{\text{S}}}{{h}_{\text{SR}}} \right),\mathcal{C}\left( {{P}_{\text{S}}}{{h}_{\text{SD}}} \right)+\mathcal{C}\left( {{P}_{\text{R}}}{{h}_{\text{RD}}} \right) \right\}}{{{P}_{\text{S}}}+{{P}_{\text{R}}}+2\alpha_\text{B}} \\
\label{r_link_trans_prb_quasi_con1}
\text{s.t. }\ &{{P}_{\text{S}}}+{{P}_{\text{R}}}\ge 2{{P}_{\text{B}}}-2\alpha_\text{B},\\
\label{r_link_trans_prb_quasi_con2}
&{{P}_{\text{S}}}\ge0,\ {{P}_{\text{R}}}\ge0,
\end{align}
where (\ref{r_link_trans_prb_quasi_con1}) is obtained by substituting $p^*$ into the constraint $0\le p\le 1$.

Recall that $h_\text{SR}\ge 2h_\text{SD}$ must be satisfied for RAT-DL.
It can be checked that
\begin{equation}
\label{min_condition}
\mathcal{C}\left( {{P}_{\text{S}}}{{h}_{\text{SR}}} \right)\ge \mathcal{C}\left( {{P}_{\text{S}}}{{h}_{\text{SD}}} \right)+\mathcal{C}\left( {{P}_{\text{R}}}{{h}_{\text{RD}}} \right);
\end{equation}
otherwise, reducing the relay transmit power and increasing the source transmit power can boost the average throughput.
Therefore, by substituting (\ref{min_condition}) into (\ref{r_link_trans_prb_quasi}), problem (\ref{r_link_trans_prb_quasi})--(\ref{r_link_trans_prb_quasi_con2}) can be rewritten as
\begin{align}
\label{r_link_trans_prb_quasi_no_min}
\mathcal{C}_\text{B}\left( {{P}_{\text{B}}} \right)=\underset{\left\{{{P}_{\text{S}}},{{P}_{\text{R}}}\right\}}{\mathop{\max }}\,&\frac{\mathcal{C}\left( {{P}_{\text{S}}}{{h}_{\text{SD}}} \right)+\mathcal{C}\left( {{P}_{\text{R}}}{{h}_{\text{RD}}} \right)}{{{P}_{\text{S}}}+{{P}_{\text{R}}}+2\alpha_\text{B}}\cdot {{P}_{\text{B}}}\\
\label{r_link_trans_prb_quasi_no_min_con1}
\text{s.t. }\ &{{P}_{\text{S}}}+{{P}_{\text{R}}}\ge 2{{P}_{\text{B}}}-2\alpha_\text{B},\\
\label{r_link_trans_prb_quasi_no_min_con2}
&{{P}_{\text{R}}}\le \frac{{{P}_{\text{S}}}\left( {{h}_{\text{SR}}}-{{h}_{\text{SD}}} \right)}{{{h}_{\text{RD}}}\left( 1+{{P}_{\text{S}}}{{h}_{\text{SD}}} \right)},\\
\label{r_link_trans_prb_quasi_no_min_con3}
&{{P}_{\text{S}}}\ge0,\ {{P}_{\text{R}}}\ge0,
\end{align}
where (\ref{r_link_trans_prb_quasi_no_min_con2}) is obtained from (\ref{min_condition}).

The characterization of the objective function (\ref{r_link_trans_prb_quasi_no_min}) is analyzed in the following lemma.
\begin{lemma}
\label{lem2_quasiconcavity}
$\frac{\mathcal{C}\left( {{P}_{\text{S}}}{{h}_{\text{SD}}} \right)+\mathcal{C}\left( {{P}_{\text{R}}}{{h}_{\text{RD}}} \right)}{{{P}_{\text{S}}}+{{P}_{\text{R}}}+2\alpha_\text{B}}$ is quasiconcave over ${{P}_{\text{S}}}\ge 0$, ${{P}_{\text{R}}}\ge 0$, and there exists a unique global maximum point.
Furthermore, it is first strictly increasing and then strictly decreasing over $P_\text{S}$ and $P_\text{R}$, respectively \cite{GMiao10}.
\end{lemma}

\subsubsection{Optimal Point}
\begin{figure}[!t]
\centering
\includegraphics[width=2in]{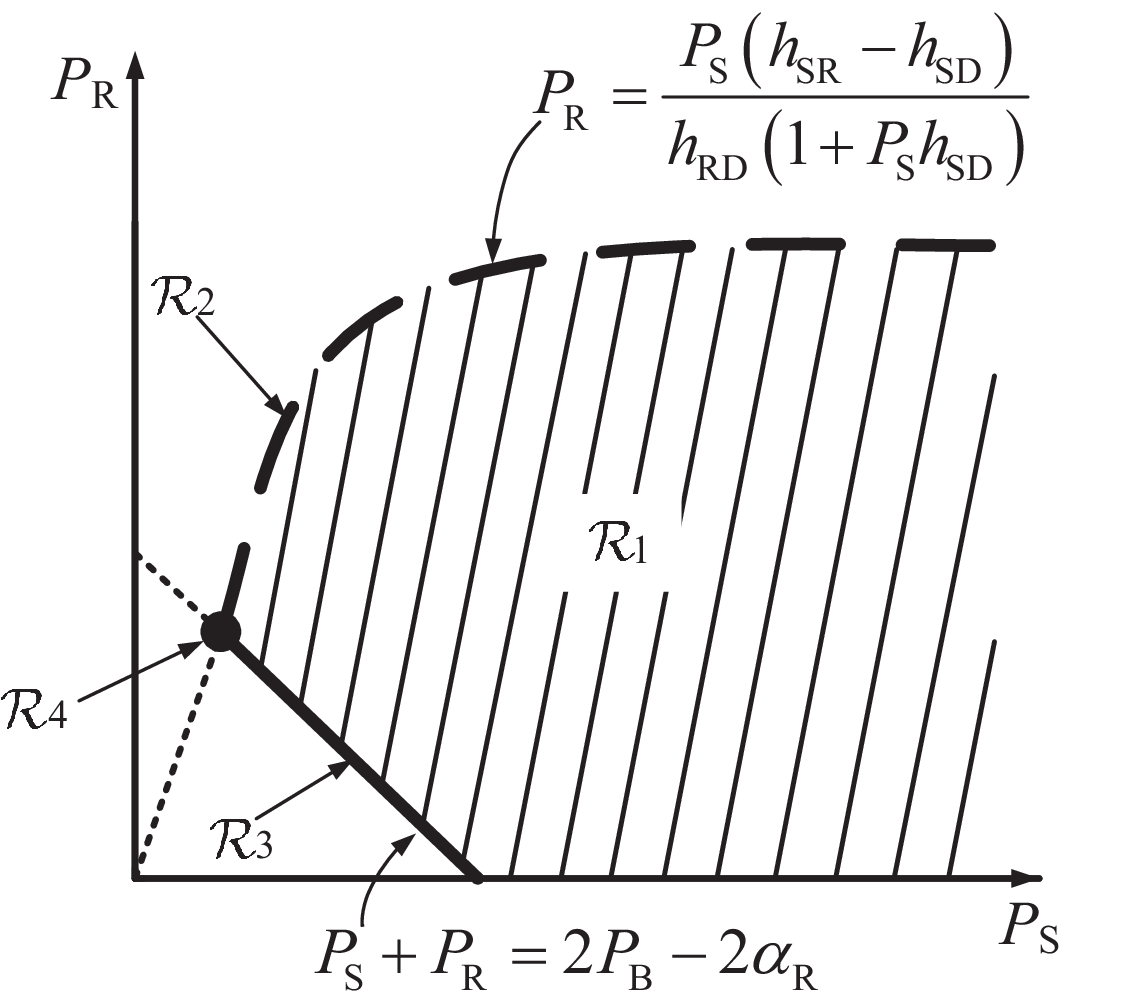}
\caption{An illustration of the four disjoint feasible subsets defined in (\ref{r_link_trans_prb_quasi_no_min_con1})--(\ref{r_link_trans_prb_quasi_no_min_con3}).}
\label{fig_fsregion}
\end{figure}
Next, the optimal point of problem (\ref{r_link_trans_prb_quasi_no_min})--(\ref{r_link_trans_prb_quasi_no_min_con3}) is derived.
By considering the combinations of the cases that the equalities in (\ref{r_link_trans_prb_quasi_no_min_con1}) and (\ref{r_link_trans_prb_quasi_no_min_con2}) are achieved or not, divide the feasible set defined by constraints (\ref{r_link_trans_prb_quasi_no_min_con1})--(\ref{r_link_trans_prb_quasi_no_min_con3}) into four disjoint parts $\left\{\mathcal{R}_i\right\}, i=1,2,3,4$, as shown in Fig. \ref{fig_fsregion}, where $\left\{\mathcal{R}_i\right\}$ are rigorously defined as follows:
\begin{align}
\nonumber
{{\mathcal{R}}_{1}}\triangleq\bigg\{& \left( {{P}_{\text{S}}},{{P}_{\text{R}}} \right)\left|{{P}_{\text{S}}}+{{P}_{\text{R}}}> 2{{P}_{\text{B}}}-2\alpha_\text{B},\right.\\
\label{feasible_set1}
&\left.\left.{{P}_{\text{R}}}< \frac{{{P}_{\text{S}}}\left( {{h}_{\text{SR}}}-{{h}_{\text{SD}}} \right)}{{{h}_{\text{RD}}}\left( 1+{{P}_{\text{S}}}{{h}_{\text{SD}}} \right)},{{P}_{\text{S}}}\ge0,{{P}_{\text{R}}}\ge0 \right. \right\},\\
\nonumber
{{\mathcal{R}}_{2}}\triangleq\bigg\{& \left( {{P}_{\text{S}}},{{P}_{\text{R}}} \right)\left|{{P}_{\text{S}}}+{{P}_{\text{R}}}> 2{{P}_{\text{B}}}-2\alpha_\text{B},\right.\\
\label{feasible_set2}
&\left.\left.{{P}_{\text{R}}}= \frac{{{P}_{\text{S}}}\left( {{h}_{\text{SR}}}-{{h}_{\text{SD}}} \right)}{{{h}_{\text{RD}}}\left( 1+{{P}_{\text{S}}}{{h}_{\text{SD}}} \right)},{{P}_{\text{S}}}\ge0,{{P}_{\text{R}}}\ge0 \right.\right\},\\
\nonumber
{{\mathcal{R}}_{3}}\triangleq\bigg\{& \left( {{P}_{\text{S}}},{{P}_{\text{R}}} \right)\left|{{P}_{\text{S}}}+{{P}_{\text{R}}}= 2{{P}_{\text{B}}}-2\alpha_\text{B},\right.\\
\label{feasible_set3}
&\left.\left.{{P}_{\text{R}}}< \frac{{{P}_{\text{S}}}\left( {{h}_{\text{SR}}}-{{h}_{\text{SD}}} \right)}{{{h}_{\text{RD}}}\left( 1+{{P}_{\text{S}}}{{h}_{\text{SD}}} \right)},{{P}_{\text{S}}}\ge0,{{P}_{\text{R}}}\ge0 \right.\right\},\\
\nonumber
{{\mathcal{R}}_{4}}\triangleq\bigg\{& \left( {{P}_{\text{S}}},{{P}_{\text{R}}} \right)\left|{{P}_{\text{S}}}+{{P}_{\text{R}}}= 2{{P}_{\text{B}}}-2\alpha_\text{B},\right.\\
\label{feasible_set4}
&\left.\left.{{P}_{\text{R}}}= \frac{{{P}_{\text{S}}}\left( {{h}_{\text{SR}}}-{{h}_{\text{SD}}} \right)}{{{h}_{\text{RD}}}\left( 1+{{P}_{\text{S}}}{{h}_{\text{SD}}} \right)},{{P}_{\text{S}}}\ge0,{{P}_{\text{R}}}\ge0 \right.\right\}.
\end{align}

Suppose that $\left(P_{\text{S}}^{*},P_{\text{R}}^{*}\right)$ is the optimal point to problem (\ref{r_link_trans_prb_quasi_no_min})--(\ref{r_link_trans_prb_quasi_no_min_con3}), this point belongs to only one \footnote{The uniqueness of the solution of problem (\ref{r_link_trans_prb_quasi_no_min})--(\ref{r_link_trans_prb_quasi_no_min_con3}) can be proved by contradiction with the property of the strictly quasiconcave function.} of the four sets defined in (\ref{feasible_set1})--(\ref{feasible_set4}).
Thus, the following four cases are studied:

\begin{enumerate}
\item Case 1: $\left(P_{\text{S}}^{*},P_{\text{R}}^{*}\right) \in {\mathcal{R}}_{1}$.

In this case, both the constraints (\ref{r_link_trans_prb_quasi_no_min_con1}) and (\ref{r_link_trans_prb_quasi_no_min_con2}) are inactive, and thus a candidate solution of problem (\ref{r_link_trans_prb_quasi_no_min})--(\ref{r_link_trans_prb_quasi_no_min_con3}) is given by maximizing its objective constraint and ignoring the constraints, i.e.,
\begin{align}
\nonumber
&\left( P_{\text{S}}^{1},P_{\text{R}}^{1} \right)=\left( {{P}_{\text{ee2}}},{{P}_{\text{ee3}}} \right)\triangleq\\
\label{r_link_trans_prb_quasi_case1_opt_candidate}
&\mathop{\arg}\underset{\left\{{{P}_{\text{S}}}\ge0,{{P}_{\text{R}}}\ge0\right\}}{\mathop{\max }}\,\frac{\mathcal{C}\left( {{P}_{\text{S}}}{{h}_{\text{SD}}} \right)+\mathcal{C}\left( {{P}_{\text{R}}}{{h}_{\text{RD}}} \right)}{{{P}_{\text{S}}}+{{P}_{\text{R}}}+2\alpha_\text{B}}.
\end{align}
After obtain $\left( {{P}_{\text{ee2}}},{{P}_{\text{ee3}}} \right)$, the feasibility condition $\left({{P}_{\text{ee2}}},{{P}_{\text{ee3}}}\right)\in \mathcal{R}_1$ needs to be doubly checked: If it is not satisfied, $\left( {{P}_{\text{ee2}}},{{P}_{\text{ee3}}} \right)$ cannot be claimed as a solution candidate for Case 1.

\item Case 2: $\left(P_{\text{S}}^{*},P_{\text{R}}^{*}\right) \in {\mathcal{R}}_{2}$.

In this case, constraint (\ref{r_link_trans_prb_quasi_no_min_con1}) is inactive and constraint (\ref{r_link_trans_prb_quasi_no_min_con2}) is active.
Since the equality in (\ref{r_link_trans_prb_quasi_no_min_con2}) is achieved, it follows ${{P}_{\text{R}}}^*= \frac{{{P}_{\text{S}}}^*\left( {{h}_{\text{SR}}}-{{h}_{\text{SD}}} \right)}{{{h}_{\text{RD}}}\left( 1+{{P}_{\text{S}}}^*{{h}_{\text{SD}}} \right)}$, with which problem (\ref{r_link_trans_prb_quasi_no_min})--(\ref{r_link_trans_prb_quasi_no_min_con3}) can be simplified as
\begin{align}
\nonumber
&\mathcal{C}_\text{B}\left( {{P}_{\text{B}}} \right)=\\
\label{r_link_trans_prb_quasi_case2}
\underset{{{P}_{\text{S}}}\ge 0}{\mathop{\max }}\;&\frac{{{P}_{\text{B}}}\left( 1+{{P}_{\text{S}}}{{h}_{\text{SD}}} \right){{h}_{\text{RD}}}\mathcal{C}\left( {{P}_{\text{S}}}{{h}_{\text{SR}}} \right)}{{{h}_{\text{SD}}}{{h}_{\text{RD}}}P_{_{\text{S}}}^{2}+\left( U+2{{h}_{\text{SD}}}{{h}_{\text{RD}}}{{P}_{\text{B}}} \right){{P}_{\text{S}}}+2\alpha_\text{B}{{h}_{\text{RD}}}} \\
\label{r_link_trans_prb_quasi_case2_con1}
\text{s.t. }\ &{{h}_{\text{SD}}}{{h}_{\text{RD}}}P_{\text{S}}^{2}+U{{P}_{\text{S}}}+2\alpha_\text{B}{{h}_{\text{RD}}}-2{{h}_{\text{RD}}}{{P}_{\text{B}}}> 0,
\end{align}
where $U$ is defined as
\begin{equation}
\label{u}
U\triangleq{{h}_{\text{SR}}}+{{h}_{\text{RD}}}-{{h}_{\text{SD}}}+2\alpha_\text{B}{{h}_{\text{SD}}}{{h}_{\text{RD}}}-2{{h}_{\text{SD}}}{{h}_{\text{RD}}}{{P}_{\text{B}}}.
\end{equation}

To solve problem (\ref{r_link_trans_prb_quasi_case2})--(\ref{r_link_trans_prb_quasi_case2_con1}), first consider the case without constraint (\ref{r_link_trans_prb_quasi_case2_con1}).
Since (\ref{r_link_trans_prb_quasi_no_min}) is quasiconcave over ${{P}_{\text{S}}}$ and ${{P}_{\text{R}}}$, and ${{P}_{\text{R}}}= \frac{{{P}_{\text{S}}}\left( {{h}_{\text{SR}}}-{{h}_{\text{SD}}} \right)}{{{h}_{\text{RD}}}\left( 1+{{P}_{\text{S}}}{{h}_{\text{SD}}} \right)}$ is nondecreasing over ${{P}_{\text{R}}}$, (\ref{r_link_trans_prb_quasi_case2}) is also quasiconcave over ${{P}_{\text{S}}}\ge0$ \cite{SBoyd04}.
Moreover, it is easy to verify that: As $P_\text{S}\to 0^+$, the objective function (\ref{r_link_trans_prb_quasi_case2}) approaches 0; as $P_\text{S}$ increases, (\ref{r_link_trans_prb_quasi_case2}) is always positive; and when $P_\text{S}\to \infty$, (\ref{r_link_trans_prb_quasi_case2}) approaches 0 again.
Therefore, it is concluded that (\ref{r_link_trans_prb_quasi_case2}) owns a global maximum point over $P_\text{S}\ge0$.
Define
\begin{align}
\nonumber
&{{P}_{\text{ee4}}}\triangleq\\
\label{def_pee4}
&\mathop{\arg}\underset{{{P}_{\text{S}}}\ge0}{\mathop{\max }}\,\frac{\left( 1+{{P}_{\text{S}}}{{h}_{\text{SD}}} \right){{h}_{\text{RD}}}\mathcal{C}\left( {{P}_{\text{S}}}{{h}_{\text{SR}}} \right)}{{{h}_{\text{SD}}}{{h}_{\text{RD}}}P_{{\text{S}}}^{2}+\left( U+2{{h}_{\text{SD}}}{{h}_{\text{RD}}}{{P}_{\text{B}}} \right){{P}_{\text{S}}}+2\alpha_\text{B}{{h}_{\text{RD}}}},
\end{align}
which achieves the maximum value of (\ref{r_link_trans_prb_quasi_case2}) without considering constraint (\ref{r_link_trans_prb_quasi_case2_con1}).

Then, doubly check the feasibility condition (\ref{r_link_trans_prb_quasi_case2_con1}): If $P_{\text{S}}={{P}_{\text{ee4}}}$ satisfies constraint (\ref{r_link_trans_prb_quasi_case2_con1}), the solution candidate of problem (\ref{r_link_trans_prb_quasi_no_min})--(\ref{r_link_trans_prb_quasi_no_min_con3}) in Case 2 is given as
\begin{equation}
\label{r_link_trans_prb_quasi_case2_opt_candidate}
\left( P_{\text{S}}^{2},P_{\text{R}}^{2} \right)=\left( P_\text{ee4},\frac{P_\text{ee4}\left( {{h}_{\text{SR}}}-{{h}_{\text{SD}}} \right)}{{{h}_{\text{RD}}}\left( 1+P_\text{ee4}{{h}_{\text{SD}}} \right)}\right),
\end{equation}
where $P_{\text{R}}^{2}=\frac{P_\text{ee4}\left( {{h}_{\text{SR}}}-{{h}_{\text{SD}}} \right)}{{{h}_{\text{RD}}}\left( 1+P_\text{ee4}{{h}_{\text{SD}}} \right)}$ is obtained with constraint (\ref{r_link_trans_prb_quasi_no_min_con2}) achieving its equality and $P_{\text{S}}^{2}=P_\text{ee4}$; and if $P_{\text{S}}^{2}={{P}_{\text{ee4}}}$ does not satisfy constraint (\ref{r_link_trans_prb_quasi_case2_con1}), $\left( P_\text{ee4},\frac{P_\text{ee4}\left( {{h}_{\text{SR}}}-{{h}_{\text{SD}}} \right)}{{{h}_{\text{RD}}}\left( 1+P_\text{ee4}{{h}_{\text{SD}}} \right)}\right)$ is not the optimal point of problem (\ref{r_link_trans_prb_quasi_no_min})--(\ref{r_link_trans_prb_quasi_no_min_con3}). Under the circumstance, the optimal solution must be on the boundary of the feasible set, i.e., constraint (\ref{r_link_trans_prb_quasi_case2_con1}) must be satisfied with equality, and this implies that the optimal point belongs to Case 4.

\item Case 3: $\left(P_{\text{S}}^{*},P_{\text{R}}^{*}\right) \in {\mathcal{R}}_{3}$.

In this case, constraint (\ref{r_link_trans_prb_quasi_no_min_con1}) is active and constraint (\ref{r_link_trans_prb_quasi_no_min_con2}) is inactive.
Since equality in (\ref{r_link_trans_prb_quasi_no_min_con1}) is achieved, it follows ${{P}_{\text{R}}}= 2{{P}_{\text{B}}}-2\alpha_\text{B}-{{P}_{\text{S}}}$, with which problem (\ref{r_link_trans_prb_quasi_no_min})--(\ref{r_link_trans_prb_quasi_no_min_con3}) can be simplified as
\begin{align}
\nonumber
\mathcal{C}_\text{B}\left( {{P}_{\text{B}}} \right)=\underset{{{P}_{\text{S}}}\ge0}{\mathop{\max }}\;&\mathcal{C}\left( -{{h}_{\text{SD}}}{{h}_{\text{RD}}}{P_{{\text{S}}}}^{2}+\left( {{h}_{\text{SR}}}-U \right){{P}_{\text{S}}}\right.\\
\label{r_link_trans_prb_quasi_case3}
&\left.+\left( 2{{P}_{\text{B}}}-2\alpha_\text{B} \right){{h}_{\text{RD}}} \right)\\
\nonumber
\text{s.t. }\ &{{h}_{\text{SD}}}{{h}_{\text{RD}}}P_{\text{S}}^{2}+U{{P}_{\text{S}}}+2\alpha_\text{B}{{h}_{\text{RD}}}\\
\label{r_link_trans_prb_quasi_case3_con1}
&-2{{h}_{\text{RD}}}{{P}_{\text{B}}}> 0.
\end{align}

As (\ref{r_link_trans_prb_quasi_case3}) is a composition of a logarithmic function and a quadratic function, the maximum point of (\ref{r_link_trans_prb_quasi_case3}) without considering constraint (\ref{r_link_trans_prb_quasi_case3_con1}) is achieved by
\begin{equation}
\label{r_link_trans_prb_quasi_case3_opt_ps}
P_\text{S}=\frac{{{h}_{\text{SD}}}-{{h}_{\text{RD}}}}{2{{h}_{\text{SD}}}{{h}_{\text{RD}}}}+{{P}_{\text{B}}}-\alpha_\text{B}.
\end{equation}
Then, doubly check the feasibility condition (\ref{r_link_trans_prb_quasi_case3_con1}): If (\ref{r_link_trans_prb_quasi_case3_opt_ps}) satisfies constraint (\ref{r_link_trans_prb_quasi_case3_con1}), the solution candidate of problem (\ref{r_link_trans_prb_quasi_no_min})--(\ref{r_link_trans_prb_quasi_no_min_con3}) in Case 3 is then given as
\begin{equation}
\label{r_link_trans_prb_quasi_case3_opt_candidate}
\left( P_{\text{S}}^{3},P_{\text{R}}^{3} \right)=\left(F,G\right),
\end{equation}
where $P_{\text{R}}^{3}=G$ is obtained with constraint (\ref{r_link_trans_prb_quasi_no_min_con1}) achieving its equality and $P_\text{S}^3=F$, and $F$, $G$ are defined as
\begin{eqnarray}
\label{f}
F\triangleq\frac{{{h}_{\text{SD}}}-{{h}_{\text{RD}}}}{2{{h}_{\text{SD}}}{{h}_{\text{RD}}}}+P_\text{B}-\alpha_\text{B},\\
\label{g}
G\triangleq\frac{{{h}_{\text{RD}}}-{{h}_{\text{SD}}}}{2{{h}_{\text{SD}}}{{h}_{\text{RD}}}}+P_\text{B}-\alpha_\text{B};
\end{eqnarray}
and if (\ref{r_link_trans_prb_quasi_case3_opt_ps}) does not satisfy constraint (\ref{r_link_trans_prb_quasi_case3_con1}), $\left(F,G\right)$ is not the optimal point of problem (\ref{r_link_trans_prb_quasi_no_min})--(\ref{r_link_trans_prb_quasi_no_min_con3}).
Under the circumstance, the optimal solution must be on the boundary of the feasible set, i.e., constraint (\ref{r_link_trans_prb_quasi_case3_con1}) must be satisfied with equality, and this implies that the optimal point belongs to Case 4.

\item Case 4: $\left(P_{\text{S}}^{*},P_{\text{R}}^{*}\right) \in {\mathcal{R}}_{4}$.

In this case, both the constraints (\ref{r_link_trans_prb_quasi_no_min_con1}) and (\ref{r_link_trans_prb_quasi_no_min_con2}) are active.
Thus, they lead to ${{P}_{\text{R}}}= \frac{{{P}_{\text{S}}}\left( {{h}_{\text{SR}}}-{{h}_{\text{SD}}} \right)}{{{h}_{\text{RD}}}\left( 1+{{P}_{\text{S}}}{{h}_{\text{SD}}} \right)}$ and ${{P}_{\text{S}}}+{{P}_{\text{R}}}= 2{{P}_{\text{B}}}-2\alpha_\text{B}$, which imply
\begin{equation}
\label{r_link_trans_prb_quasi_case4}
{{h}_{\text{SD}}}{{h}_{\text{RD}}}P_{\text{S}}^{2}+U{{P}_{\text{S}}}-2\left({{P}_{\text{B}}}-\alpha_\text{B}\right){{h}_{\text{RD}}}= 0,
\end{equation}
where $U$ is defined in (\ref{u}).
The solution can be readily obtained within $P_{\text{S}}\ge0$.
Denote $V$ as the positive solution of (\ref{r_link_trans_prb_quasi_case4}), i.e.,
\begin{equation}
\label{v}
V\triangleq\frac{-U+\sqrt{{{U}^{2}}+8\left({{P}_{\text{B}}}-\alpha_\text{B}\right){{h}_{\text{SD}}}{{h}_{\text{RD}}}^2}}{2{{h}_{\text{SD}}}{{h}_{\text{RD}}}}.
\end{equation}
The solution candidate of problem (\ref{r_link_trans_prb_quasi_no_min})--(\ref{r_link_trans_prb_quasi_no_min_con3}) in Case 4 is given as
\begin{equation}
\label{r_link_trans_prb_quasi_case4_opt_candidate}
\left( P_{\text{S}}^{4},P_{\text{R}}^{4} \right)=\left( V,2{{P}_{\text{B}}}-2\alpha_\text{B}-V\right),
\end{equation}
where $P_{\text{R}}^{4}=2{{P}_{\text{B}}}-2\alpha_\text{B}-V$ is obtained with constraint (\ref{r_link_trans_prb_quasi_no_min_con1}) reaching equality and $P_{\text{S}}^{4}=V$.
\end{enumerate}

\begin{remark}
\label{rmk_optimal_sol}
After problem (\ref{r_link_trans_prb_quasi_no_min})--(\ref{r_link_trans_prb_quasi_no_min_con3}) is solved under the above four cases, the one achieves the largest optimal value among the four solution candidates is the optimal solution of problem (\ref{r_link_trans_prb_quasi_no_min})--(\ref{r_link_trans_prb_quasi_no_min_con3}).
\end{remark}

\subsubsection{Optimal Value}
With four optimal solution candidates obtained, the corresponding necessary and sufficient conditions that allow each case to happen are studied, and the average throughput for RAT-DL is derived.
\begin{enumerate}
\item If the candidate solution $\left({{P}_{\text{ee2}}},{{P}_{\text{ee3}}}\right)$ obtained in (\ref{r_link_trans_prb_quasi_case1_opt_candidate}) is the solution to problem (\ref{r_link_trans_prb_quasi_no_min})--(\ref{r_link_trans_prb_quasi_no_min_con3}), according to Lemma \ref{lem2_quasiconcavity} and Remark \ref{rmk_optimal_sol}, the necessary and sufficient condition that Case 1 happens is given as ${{S}_{1}}\cap {{S}_{2}}$, where
\begin{equation}
{{S}_{1}}\triangleq\left\{ {{P}_{\text{B}}}\left|{{P}_{\text{ee2}}}+{{P}_{\text{ee3}}}> 2{{P}_{\text{B}}}-2\alpha_\text{B} \right.\right\},
\end{equation}
\begin{equation}
{{S}_{2}}\triangleq\begin{cases}
\mathbb{R}^{+} & \text{if }{{P}_{\text{ee3}}}< \frac{{{P}_{\text{ee2}}}\left( {{h}_{\text{SR}}}-{{h}_{\text{SD}}} \right)}{{{h}_{\text{RD}}}\left( 1+{{P}_{\text{ee2}}}{{h}_{\text{SD}}} \right)}\\
\emptyset & \text{otherwise.}
\end{cases}
\end{equation}

\item If the candidate solution $\left( P_\text{ee4},\frac{P_\text{ee4}\left( {{h}_{\text{SR}}}-{{h}_{\text{SD}}} \right)}{{{h}_{\text{RD}}}\left( 1+P_\text{ee4}{{h}_{\text{SD}}} \right)}\right)$ obtained in (\ref{r_link_trans_prb_quasi_case2_opt_candidate}) is the solution to problem (\ref{r_link_trans_prb_quasi_no_min})--(\ref{r_link_trans_prb_quasi_no_min_con3}), according to Lemma \ref{lem2_quasiconcavity} and Remark \ref{rmk_optimal_sol}, the necessary and sufficient condition that Case 2 happens is given as $S^c_2\cap {{S}_{3}}$, where $S^c_2$ is the complementary set of $S_2$ and
\begin{equation}
{{S}_{3}}\triangleq\left\{ {{P}_{\text{B}}}\left|{{P}_{\text{ee4}}}+\frac{{{P}_{\text{ee4}}}\left( {{h}_{\text{SR}}}-{{h}_{\text{SD}}} \right)}{{{h}_{\text{RD}}}\left( 1+{{P}_{\text{ee4}}}{{h}_{\text{SD}}} \right)}> 2{{P}_{\text{B}}}-2\alpha_\text{B} \right.\right\}.
\end{equation}

\item If the candidate solution $\left( F,G\right)$ obtained in (\ref{r_link_trans_prb_quasi_case3_opt_candidate}) is the solution to problem (\ref{r_link_trans_prb_quasi_no_min})--(\ref{r_link_trans_prb_quasi_no_min_con3}), according to Lemma \ref{lem2_quasiconcavity} and Remark \ref{rmk_optimal_sol}, the necessary and sufficient condition that Case 3 happens is given as $S^c_1\cap {{S}_{2}} \cap S_4$, where $S^c_1$ is the complementary set of $S_1$ and
\begin{equation}
{{S}_{4}}\triangleq\left\{ {{P}_{\text{B}}}\left|G <\frac{F\left( {{h}_{\text{SR}}}-{{h}_{\text{SD}}} \right)}{{{h}_{\text{RD}}}\left( 1+F{{h}_{\text{SD}}} \right)}\right. \right\}.
\end{equation}

\item If the candidate solution $\left( V,2{{P}_{\text{B}}}-2\alpha_\text{B}-V\right)$ obtained in (\ref{r_link_trans_prb_quasi_case4_opt_candidate}) is the solution to problem (\ref{r_link_trans_prb_quasi_no_min})--(\ref{r_link_trans_prb_quasi_no_min_con3}), according to Lemma \ref{lem2_quasiconcavity} and Remark \ref{rmk_optimal_sol}, the necessary and sufficient condition that Case 4 happens is given as $\left(S_1^c\cap S_2\cap S_4^c\right)\cup \left(S_2^c\cap S_3^c\right)$, where $S^c_3$ and $S^c_4$ are the complementary sets of $S_3$ and $S_4$, respectively.
\end{enumerate}

Based on the above discussions, the optimal solutions of problem (\ref{r_link_origin_prb}) are summarized.
The average throughput defined in (\ref{r_link_origin_prb}) for RAT-DL is also given in the following proposition.
\begin{proposition}
\label{prp2_CRPB_char}
The optimal power allocation for problem (\ref{r_link_origin_prb}) is given as: Transmit with power value $\left( P_{\text{S}}^{*},P_{\text{R}}^{*} \right)$ over $p^*$ portion of time slots and keep silent for the rest of slots, where
\begin{equation}
\label{optimal_sol_CRPB}
\left( P_{\text{S}}^{*},P_{\text{R}}^{*} \right)= \begin{cases}
   \left( {{P}_{\text{ee2}}},{{P}_{\text{ee3}}} \right) & P_\text{B}\in {{S}_{1}}\cap {{S}_{2}}  \\
   \left( {{P}_{\text{ee4}}},\frac{{{P}_{\text{ee4}}}\left( {{h}_{\text{SR}}}-{{h}_{\text{SD}}} \right)}{{{h}_{\text{RD}}}\left( 1+{{P}_{\text{ee4}}}{{h}_{\text{SD}}} \right)} \right) & P_\text{B}\in S_{_{2}}^{c}\cap {{S}_{3}}  \\
   \left(F,G \right) & P_\text{B}\in S_{1}^{c}\cap {{S}_{2}}\cap {{S}_{4}}  \\
   \left( V,2{{P}_{\text{B}}}-2\alpha_\text{B}-V\right) & \text{otherwise},
\end{cases}
\end{equation}
and $p^*=\frac{2{{P}_{\text{B}}}}{{{P}_{\text{S}}^{*}}+{{P}_{\text{R}}^{*}}+2\alpha_\text{B}}$.
With the optimal power allocation, the average throughput $\mathcal{C}_\text{B}\left( {{P}_{\text{B}}} \right)$ defined in (\ref{r_link_origin_prb}) is given as
\begin{equation}
\label{crpb}
\mathcal{C}_\text{B}\left( {{P}_{\text{B}}} \right)= \begin{cases}
   J {{P}_{\text{B}}} & P_\text{B}\in{{S}_{1}}\cap {{S}_{2}}  \\
   M {{P}_{\text{B}}} & P_\text{B}\in S_{_{2}}^{c}\cap {{S}_{3}}  \\
   \frac{1}{2}\mathcal{C}\left( Fh_\text{SD} \right)+\frac{1}{2}\mathcal{C}\left( Gh_\text{RD}\right) & P_\text{B}\in  S_{1}^{c}\cap {{S}_{2}}\cap {{S}_{4}}  \\
   \frac{1}{2}\mathcal{C}\left( V{{h}_{\text{SR}}} \right) & \text{otherwise},  \\
\end{cases}
\end{equation}
which is continuous, differentiable, and concave over the domain, where $J\triangleq\frac{\mathcal{C}\left( {{P}_{\text{ee2}}}{{h}_{\text{SD}}} \right)+\mathcal{C}\left( {{P}_{\text{ee3}}}{{h}_{\text{RD}}} \right) }{{{P}_{\text{ee2}}}+{{P}_{\text{ee3}}}+2\alpha_\text{B}}$ and $M\triangleq\frac{\left( 1+{{P}_{\text{ee4}}}{{h}_{\text{SD}}} \right){{h}_{\text{RD}}}\mathcal{C}\left( {{P}_{\text{ee4}}}{{h}_{\text{SR}}} \right)}{{{h}_{\text{SD}}}{{h}_{\text{RD}}}P_{_{\text{ee4}}}^{2}+\left( U+2{{h}_{\text{SD}}}{{h}_{\text{RD}}}{{P}_{\text{B}}}  \right){{P}_{\text{ee4}}}+2\alpha_\text{B}{{h}_{\text{RD}}}}$.
\end{proposition}
\begin{IEEEproof}
Please see Appendix \ref{prf_prp2_CRPB_char}.
\end{IEEEproof}

\begin{remark}
\label{rmk2_CRPB_char}
Based on Proposition \ref{prp2_CRPB_char}, it is worth noting that the transmission scheme given in Proposition \ref{prp2_CRPB_char} is similar to DLT, which transmits with an on-off structure when the average power budget $P_\text{B}$ is small to maximize the EE for the case of RAT-DL, and transmits constantly when the average power budget $P_\text{B}$ is large to maximize the SE.
It is also worth noticing that $P_\text{ee2}$, $P_\text{ee3}$, and $P_\text{ee4}$ can be efficiently obtained by a simple bisection search.
\end{remark}

\section{Optimal Power Allocation for the Case with Direct Link}
\label{sec_opt}
Based on the two special scenarios studied in the previous section, the optimal power allocation for the case with direct link between the source and the destination is investigated.

\subsection{Optimal Power Allocation for the Mixed Transmission}
\label{subsec_mt}
It is worth pointing out that the considered system can only work in one of three modes for each time slot: DLT, RAT-DL, or keeping silence.
Thus, the optimal power allocation over an infinite time horizon can only be the combination of the above three modes.
Moreover, by the analysis in the previous section, it is shown that the mode of keeping silence can be incorporated into any one of the first two modes, since there is no transmission in portion of the time slots in these two modes.
Therefore, the optimal power allocation for the throughput maximization of the relay channel with direct link is a mixed transmission (MT) scheme, i.e., transmit with the schemes of DLT and RAT-DL discussed in the previous section.
In other words, to solve the optimal power allocation for MT is equivalent to find the average power budgets $P_\text{A}$ and $P_\text{B}$ for DLT and RAT-DL to maximize the throughput of the considered relay system, subject to the average power constraint $P_0$.
Then, the following proposition is easily obtained.

\begin{proposition}
The throughput maximization problem for the relay channel with direct link and non-ideal circuit power is formulated as
\begin{align}
\label{m_link_origin_prb}
\underset{\left\{{{P}_{\text{A}}},{{P}_{\text{B}}},\theta \right\}}{\mathop{\max }}\,&\theta \mathcal{C}_\text{A}\left( {{P}_{\text{A}}} \right)+\left( 1-\theta  \right)\mathcal{C}_\text{B}\left( {{P}_{\text{B}}} \right)\\
\label{m_link_origin_prb_con1}
\text{s.t. }\;\; &\theta {{P}_{\text{A}}}+\left( 1-\theta  \right){{P}_{\text{B}}}={{P}_{0}},\\
\label{m_link_origin_prb_con2}
&0\le \theta \le 1,\ {{P}_{\text{A}}}\ge 0,\ {{P}_{\text{B}}}\ge 0,
\end{align}
where $\theta$ stands for $\theta$ portion of time slots for DLT and $1-\theta$ stands for $ 1-\theta $ portion of time slots for RAT-DL.
\end{proposition}

Before the optimal solutions of problem (\ref{m_link_origin_prb})--(\ref{m_link_origin_prb_con2}) are given, the relationship between the average throughput for DLT $\mathcal{C}_\text{A}\left(P\right)$ and RAT-DL $\mathcal{C}_\text{B}\left(P\right)$ is discussed.
From (\ref{CDPA}) and (\ref{crpb}), it can be inferred that $\mathcal{C}_\text{A}\left(P\right)$ and $\mathcal{C}_\text{B}\left(P\right)$ are both increasing and concave functions, which start from the origin point, increase linearly, and then turn to logarithmic functions after some points.
Their relationship falls into the following three cases (the categorization is discussed later in Remark \ref{rmk_mt_low_snr} and Remark \ref{rmk_mt_high_snr}):
\begin{enumerate}
\item Case 1: the linear parts of $\mathcal{C}_\text{A}\left(P\right)$ and $\mathcal{C}_\text{B}\left(P\right)$ coincide, and $\mathcal{C}_\text{A}\left(P\right)> \mathcal{C}_\text{B}\left(P\right)$ after a specific point.
\item Case 2: $\mathcal{C}_\text{A}\left(P\right)> \mathcal{C}_\text{B}\left(P\right)$ for any $P>0$, i.e., $\mathcal{C}_\text{A}\left(P\right)> \mathcal{C}_\text{B}\left(P\right)$ have no intersection point for $P>0$.
\item Case 3: $\mathcal{C}_\text{A}\left(P\right)$ and $\mathcal{C}_\text{B}\left(P\right)$ have one or more intersection points for $P>0$, and $\mathcal{C}_\text{A}\left(P\right)>\mathcal{C}_\text{B}\left(P\right)$ when $P$ is large enough.
\end{enumerate}

\begin{figure}[!t]
\centering
\subfloat[]{\includegraphics[width=2in]{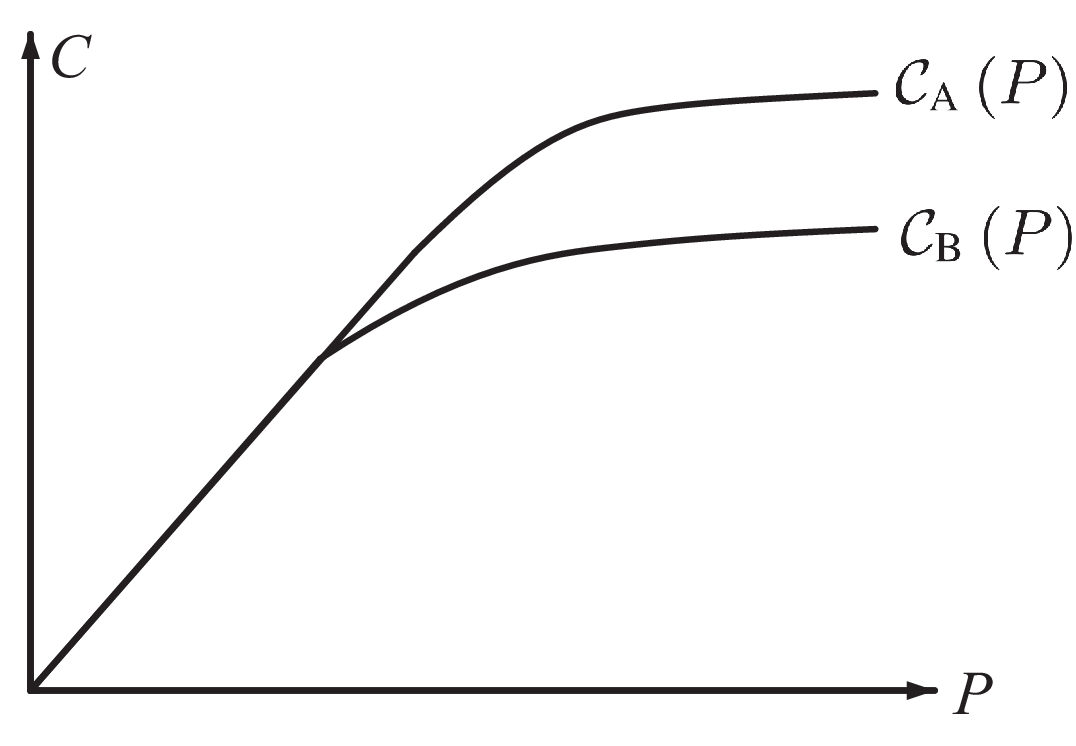}
\label{mixcase0}}
\hfil
\subfloat[]{\includegraphics[width=2in]{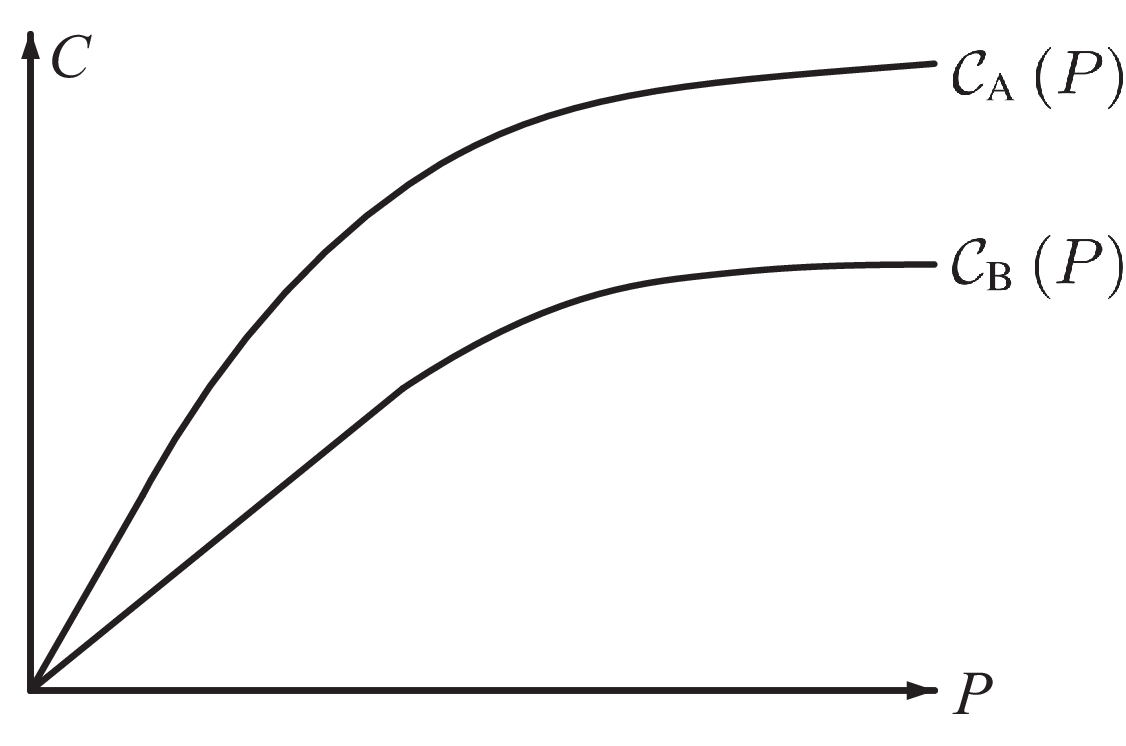}
\label{mixcase1}}
\hfil
\subfloat[]{\includegraphics[width=2in]{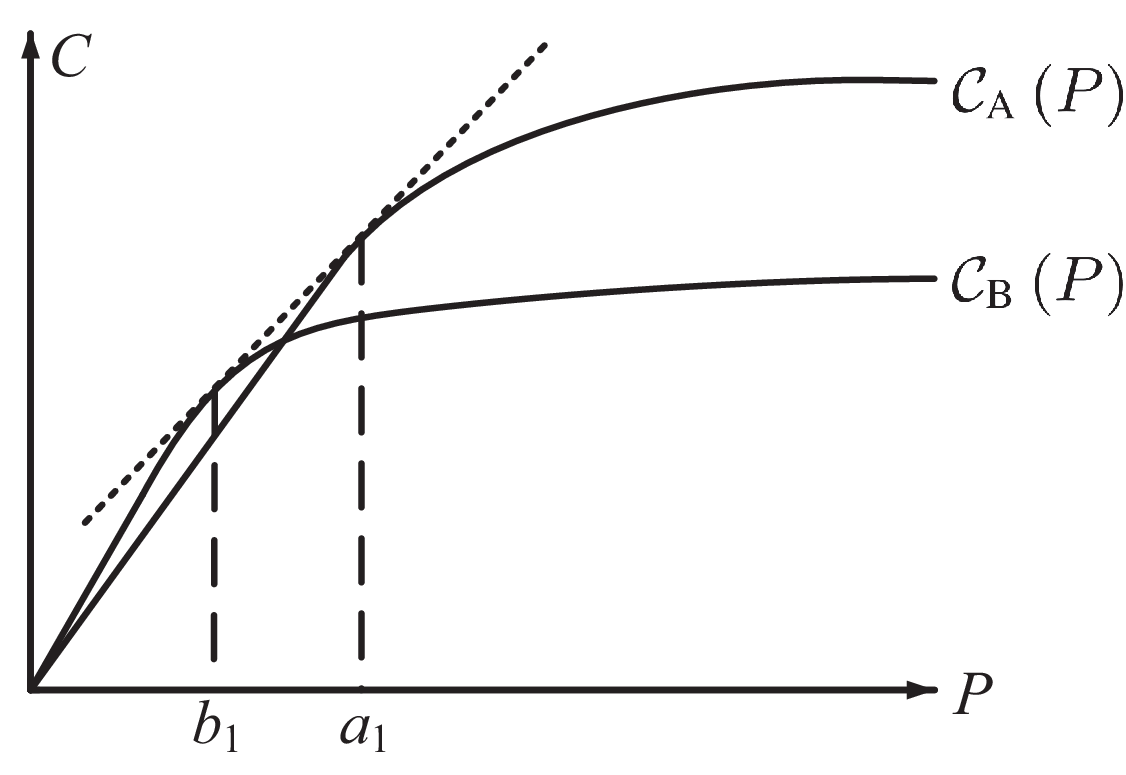}
\label{mixcase2}}
\hfil
\subfloat[]{\includegraphics[width=2in]{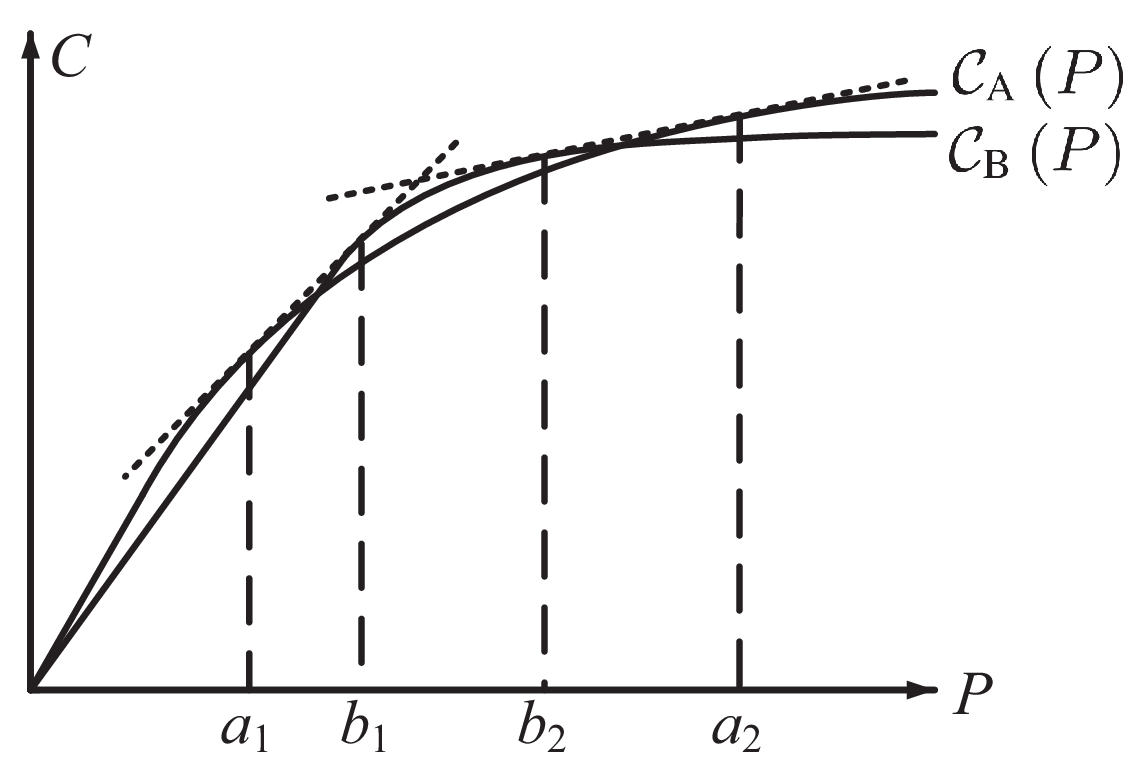}
\label{mixcase3}}
\caption{Examples of relationships between $\mathcal{C}_\text{A}\left(P\right)$ and $\mathcal{C}_\text{B}\left(P\right)$: (a) the linear parts of $\mathcal{C}_\text{A}\left(P\right)$ and $\mathcal{C}_\text{B}\left(P\right)$ coincide; (b) $\mathcal{C}_\text{A}\left(P\right)> \mathcal{C}_\text{B}\left(P\right)$ for any $P>0$; (c) $\mathcal{C}_\text{A}\left(P\right)$ and $\mathcal{C}_\text{B}\left(P\right)$ have only one intersection point for $P>0$; (d) $\mathcal{C}_\text{A}\left(P\right)$ and $\mathcal{C}_\text{B}\left(P\right)$ have two intersection points for $P>0$.}
\label{fig_mixcase}
\end{figure}

In Case 3, suppose there are $K>0$ intersection points, and there exist $K$ straight lines tangent to both $\mathcal{C}_\text{A}\left(P\right)$ and $\mathcal{C}_\text{B}\left(P\right)$.
Denote the x-coordinates of the tangent points on $\mathcal{C}_\text{A}\left(P\right)$ and $\mathcal{C}_\text{B}\left(P\right)$ as $a_i$ and $b_i$, $i=1,2,\cdot\cdot\cdot,K$, respectively.
The relationship between $a_i$ and $b_i$ is $b_1<a_1<a_2<b_2<\cdot\cdot\cdot<b_K<a_K$ if $K$ is odd, or $a_1<b_1<b_2<a_2<\cdot\cdot\cdot<b_K<a_K$ if $K$ is even.
Examples are shown in Fig. \ref{fig_mixcase}.
It is worth noting that with the average throughput $\mathcal{C}_\text{A}\left( {{P}_{\text{A}}} \right)$ and $\mathcal{C}_\text{B}\left( {{P}_{\text{B}}} \right)$ given in (\ref{CDPA}) and (\ref{crpb}), the x-coordinates $a_i$ and $b_i$ of the tangent points can be obtained by the following lemma.
\begin{lemma}
\label{lem_tangentpoints}
The x-coordinates $a$ and $b$ of the the tangent points $\left(a,\mathcal{C}_\text{A}\left( a \right)\right)$ and $\left(b,\mathcal{C}_\text{B}\left( b \right)\right)$ on the same tangent line can be obtained by solving the following two equations
\begin{eqnarray}
&&\mathcal{C}^{'}_\text{A}\left( a \right)=\mathcal{C}^{'}_\text{B}\left( b \right),\\
&&\mathcal{C}^{'}_\text{A}\left( a \right)=\frac{\mathcal{C}_\text{A}\left( a \right)-\mathcal{C}_\text{B}\left( b \right)}{a-b}.
\end{eqnarray}
If there are infinite solutions, Case 1 satisfies.
If there is no solution, Case 2 satisfies.
If there are finite solutions, Case 3 satisfies.
\end{lemma}

With the x-coordinates of the tangent points obtained by Lemma \ref{lem_tangentpoints}, the optimal power allocation for problem (\ref{m_link_origin_prb})--(\ref{m_link_origin_prb_con2}) is given as the following proposition.
\begin{proposition}
\label{prp3_mix_opt_sol}
The optimal power allocation for problem (\ref{m_link_origin_prb})--(\ref{m_link_origin_prb_con2}) is given as
\begin{enumerate}
\item For Case 1 and Case 2,
\begin{equation}
\left( P_{\text{A}}^{*},P_{\text{B}}^{*},\theta^{*} \right)=\left( {{P}_{0}},0,1 \right),
\end{equation}
\item For Case 3, if $K$ is odd,
\begin{equation}
\left( P_{\text{A}}^{*},P_{\text{B}}^{*},\theta^{*} \right)= \begin{cases}
   \left( {{P}_{0}},0,1 \right) &  P_0\in   \Omega_1\\
   \left( 0,{{P}_{0}},0 \right) &  P_0\in   \Omega_2 \\
   \left( a_i,b_i,\frac{P_0-b_i}{a_i-b_i} \right) & P_0\in \Omega_3,  \\
\end{cases}
\end{equation}
if $K$ is even,
\begin{equation}
\left( P_{\text{A}}^{*},P_{\text{B}}^{*},\theta^{*} \right)= \begin{cases}
   \left( {{P}_{0}},0,1 \right) &  P_0\in   \Omega_4\\
   \left( 0,{{P}_{0}},0 \right) &  P_0\in   \Omega_5 \\
   \left( a_i,b_i,\frac{P_0-b_i}{a_i-b_i} \right) & P_0\in \Omega_6,  \\
\end{cases}
\end{equation}
where $a_0$, $b_0$, and $a_{K+1}$ are defined as $a_0=b_0\triangleq0$ and $a_{K+1}\triangleq+\infty$ for the purpose of exposition, and $\Omega_i,i=1,2,\cdot\cdot\cdot,6$ are defined as
\begin{align}
\nonumber
&\Omega_1\triangleq \bigcup\limits_{i=1}^{\left(K+1\right)/2}\left(a_{2i-1},a_{2i}\right),\ \Omega_2\triangleq \bigcup\limits_{i=1}^{\left(K+1\right)/2}\left(b_{2i-2},b_{2i-1}\right),\\
\nonumber
&\Omega_3\triangleq \bigcup\limits_{i=1}^{\left(K-1\right)/2}\left(a_{2i},b_{2i}\right)\cup\bigcup\limits_{i=1}^{\left(K+1\right)/2}\left(b_{2i-1},a_{2i-1}\right),\\
\nonumber
&\Omega_4\triangleq \bigcup\limits_{i=0}^{K/2}\left(a_{2i},a_{2i+1}\right),\ \Omega_5\triangleq \bigcup\limits_{i=1}^{K/2}\left(b_{2i-1},b_{2i}\right),\\
\nonumber
&\Omega_6\triangleq \bigcup\limits_{i=1}^{K/2}\left(a_{2i-1},b_{2i-1}\right)\cup\bigcup\limits_{i=1}^{K/2}\left(b_{2i},a_{2i}\right).
\end{align}
\end{enumerate}
\end{proposition}
\begin{IEEEproof}
Please see Appendix \ref{prf_prp3_mix_opt_sol}.
\end{IEEEproof}

\begin{remark}
It is observed that in Case 1, Case 2, and some situations in Case 3, the optimal power allocation scheme only chooses DLT or RAT-DL, while in other situations in Case 3, a time sharing of both transmissions is applied.
The transmission types, on-off transmission or constant transmission, are decided according to DLT average power budget $P^*_\text{A}$ and RAT-DL average power budget $P^*_\text{B}$, respectively.
\end{remark}

\subsection{Asymptotic Analysis}
\label{subsec_mt_aa}
In this subsection, throughput performances for DLT and RAT-DL at the low SNR and high SNR regimes are investigated to further illustrate the optimal transmission scheme.

\subsubsection{Low SNR Regime}
\label{subsec_lowsnr}
As ${{P}_{\text{A}}}\to 0$ and ${{P}_{\text{B}}}\to 0$, the average throughput for DLT and RAT-DL at the low SNR regime are given in (\ref{CDPA}) and (\ref{crpb}):
\begin{equation}
\label{CDPA_low_SNR}
\mathcal{C}_\text{A}\left( {{P}_{\text{A}}} \right)=\frac{{\mathcal{C}}\left( {{P}_{\text{ee1}}}{{h}_{\text{SD}}} \right)}{{{P}_{\text{ee1}}}+\alpha_\text{A}} \cdot {{P}_{\text{A}}},
\end{equation}
\begin{equation}
\label{CRPB_low_SNR}
\mathcal{C}_\text{B}\left( {{P}_{\text{B}}} \right)= \begin{cases}
   \frac{ {\mathcal{C}}\left( {{P}_{\text{ee2}}}{{h}_{\text{SD}}} \right)+{\mathcal{C}}\left( {{P}_{\text{ee3}}}{{h}_{\text{RD}}}  \right)}{{{P}_{\text{ee2}}}+{{P}_{\text{ee3}}}+2\alpha_\text{B}}\cdot {{P}_{\text{B}}} & {{S}_{2}}  \\
   \frac{\left( 1+{{P}_{\text{ee4}}}{{h}_{\text{SD}}} \right){{h}_{\text{RD}}}{\mathcal{C}}\left( {{P}_{\text{ee4}}}{{h}_{\text{SR}}} \right)}{{{h}_{\text{SD}}}{{h}_{\text{RD}}}P_{_{\text{ee4}}}^{2}+\left( U+2{{h}_{\text{SD}}}{{h}_{\text{RD}}}{{P}_{\text{B}}}  \right){{P}_{\text{ee4}}}+2\alpha_\text{B}{{h}_{\text{RD}}}}\cdot {{P}_{\text{B}}} &  S_{_{2}}^{c}.  \\
\end{cases}
\end{equation}

It is interesting to note that both $\mathcal{C}_\text{A}\left( {{P}_{\text{A}}} \right)$ and $\mathcal{C}_\text{B}\left( {{P}_{\text{B}}} \right)$ are linear functions of the average power budgets ${{P}_{\text{A}}}$ and ${{P}_{\text{B}}}$ respectively at the low SNR regime.
The scaling factors $\frac{{\mathcal{C}}\left( {{P}_{\text{ee1}}}{{h}_{\text{SD}}} \right)}{{{P}_{\text{ee1}}}+\alpha_\text{A}}$ and $\left(\frac{ {\mathcal{C}}\left( {{P}_{\text{ee2}}}{{h}_{\text{SD}}} \right)+{\mathcal{C}}\left( {{P}_{\text{ee3}}}{{h}_{\text{RD}}} \right) }{{{P}_{\text{ee2}}}+{{P}_{\text{ee3}}}+2\alpha_\text{B}}\; \text{or}\; \frac{\left( 1+{{P}_{\text{ee4}}}{{h}_{\text{SD}}} \right){{h}_{\text{RD}}}{\mathcal{C}}\left( {{P}_{\text{ee4}}}{{h}_{\text{SR}}} \right)}{{{h}_{\text{SD}}}{{h}_{\text{RD}}}P_{_{\text{ee4}}}^{2}+\left( U+2{{h}_{\text{SD}}}{{h}_{\text{RD}}}{{P}_{\text{B}}}  \right){{P}_{\text{ee4}}}+2\alpha_\text{B}{{h}_{\text{RD}}}}\right)$ are the maximum EE for the case of DLT and RAT-DL, respectively.
\begin{remark}
\label{rmk_mt_low_snr}
The optimal transmission for MT chooses the one with higher EE to transmit when ${P}_0$ is small, where the corresponding EE are the scaling factors of (\ref{CDPA_low_SNR}) and (\ref{CRPB_low_SNR}).
\end{remark}

\subsubsection{High SNR Regime}
Based on the results in (\ref{CDPA}) and (\ref{crpb}), as ${{P}_{\text{A}}}\to \infty $ and ${{P}_{\text{B}}}\to \infty$, the average throughput for DLT and RAT-DL at the high SNR regime are asymptotically given as
\begin{equation}
\mathcal{C}_\text{A}\left( {{P}_{\text{A}}} \right)\approx{{\log }_{2}}\left( \left( {{P}_{\text{A}}}-\alpha_\text{A} \right){{h}_{\text{SD}}} \right),
\end{equation}
\begin{equation}
\label{CRPBHSNR}
\mathcal{C}_\text{B}\left( {{P}_{\text{B}}} \right)\approx\frac{1}{2}{{\log }_{2}}\left( V{{h}_{\text{SR}}} \right).
\end{equation}

Note that $V$ defined in (\ref{v}) is a polynomial of ${{P}_{\text{B}}}$ with maximum exponent of 1.
Besides, it is obviously obtained in (\ref{CRPBHSNR}) that the multiplexing gain of RAT-DL is $\frac{1}{2}$, which is due to the half-duplex penalty.
However, the power gain of RAT-DL does not reach the square of ${{P}_{\text{B}}}$ and can not compensate the loss in multiplexing gain, which results in a lower throughput performance at the high SNR regime compared to that of DLT.
It is also in accordance with the observed four cases of relationships between $\mathcal{C}_\text{A}\left(P\right)$ and $\mathcal{C}_\text{B}\left(P\right)$ in Section \ref{sec_opt}.
\begin{remark}
\label{rmk_mt_high_snr}
The optimal transmission for MT always chooses DLT when ${P}_0$ is large due to its higher throughput performance at the high SNR regime.
\end{remark}

\section{Relay Assisted Transmission without Direct Link}
\label{sec_relaywod}
In this section, the optimal power allocation and throughput performance of RAT-WDL are studied as a comparison.

\subsection{Optimal Power Allocation for RAT-WDL}
\label{subsec_rt}
In this subsection, the optimal power allocation and average throughput for RAT-WDL are obtained.
For RAT-WDL, the direct link between the source and the destination is inactive, i.e., the destination can only receive signals from the relay.
With the signal model described in Section \ref{sec_sys_mdl_signal_model} without considering the direct link, the transmission rate for RAT-WDL at time slot $i$ is given as
\begin{equation}
\label{rate_c}
{{R}}_\text{C}\left( i \right)=\frac {1}{2}\min \left\{  \mathcal{C}\left( {{P}_{\text{S}}}\left( i \right){{h}_{\text{SR}}} \right), \mathcal{C}\left({{P}_{\text{R}}}\left( i \right){{h}_{\text{RD}}} \right) \right\}.
\end{equation}
Denote ${\alpha_\text{C}}$ as the total circuit power consumption in the active mode for RAT-WDL, and it is the same as $\alpha_\text{B}$ in (\ref{alpha_r}) without considering $\frac{1}{2}P_\text{cr}^\text{D}$ in the first parentheses.
The total power consumption $P^\text{C}_\text{total}\left( i \right)$ at time slot $i$ is the same as $P^\text{B}_\text{total}\left( i \right)$ in (\ref{P_total^R}) with $\alpha_\text{B}$ replaced by ${\alpha_\text{C}}$.
Then, the average total power consumption for RAT-WDL is defined over $N$ time slots, as $N$ goes to infinity, i.e.,
\begin{equation}
\label{rwod_total_power_cons}
\underset{N\to \infty }{\mathop{\lim }}\,\frac{1}{N}\sum\limits_{i=1}^{N}P^\text{C}_\text{total}\left(i\right)\le P_\text{C},
\end{equation}
where $P_\text{C}\ge 0$ is the power budget.

The goal is to determine $\left\{ P_\text{S}\left( i \right) \right\}$ and $\left\{{{P}_{\text{R}}}\left( i \right)\right\}$ such that the long term average throughput subject to the average power constraint defined in (\ref{rwod_total_power_cons}) is maximized over $N$ time slots as $N\to \infty$, i.e., solve the following optimization problem
\begin{align}
\label{rwod_origin_prb}
\mathcal{C}_\text{C}\left( {{P}_{\text{C}}} \right)=\underset{\left\{{{P}_{\text{S}}}\left( i \right)\right\},\left\{{{P}_{\text{R}}}\left( i \right)\right\}}{\mathop{\max }}\,&\underset{N\to \infty }{\mathop{\lim }}\,\frac{1}{N}\sum\limits_{i=1}^{N}R_\text{C}\left( i \right)\\
\label{rwod_origin_prb_cons}
\text{s.t. }\;\;\;\;\;\; &(\ref{rwod_total_power_cons}),\ {{P}_{\text{S}}}\left( i \right)\ge 0,\ {{P}_{\text{R}}}\left( i \right)\ge 0.
\end{align}

Since objective function (\ref{rwod_origin_prb}) is nonnegative and concave, it is easy to check \cite{JXu14} that the optimal power allocation of problem (\ref{rwod_origin_prb})--(\ref{rwod_origin_prb_cons}) is given as: Transmit with power ${{P}_{\text{S}}}\left( i \right)={{P}_{\text{S}}}>0$ and ${{P}_{\text{R}}}\left( i \right)={{P}_{\text{R}}}>0$ over $p$ portion of time slots and keep silent for the rest of the slots, where $P_\text{S}$ and $P_\text{R}$ are constants.
As a result, problem (\ref{rwod_origin_prb})--(\ref{rwod_origin_prb_cons}) can be reformulated as
\begin{align}
\label{rwod_link_trans_prb}
\mathcal{C}_\text{C}\left( {{P}_{\text{C}}} \right)=\underset{\left\{{{P}_{\text{S}}},{{P}_{\text{R}}},p\right\}}{\mathop{\max }}\ &\frac{p}{2}\min \left\{ \mathcal{C}\left( {{P}_{\text{S}}}{{h}_{\text{SR}}} \right),\mathcal{C}\left( {{P}_{\text{R}}}{{h}_{\text{RD}}} \right) \right\}\\
\label{rwod_link_trans_prb_con1}
\text{s.t. }\;\;\; &\left( \frac{1}{2}{{P}_{\text{S}}}+\frac{1}{2}{{P}_{\text{R}}}+{\alpha_\text{C}} \right)\cdot p\le {{P}_{\text{C}}},\\
\label{rwod_link_trans_prb_con2}
&0\le p\le 1,\ {{P}_{\text{S}}}\ge0,\ {{P}_{\text{R}}}\ge0,
\end{align}
where (\ref{rwod_link_trans_prb_con1}) is obtained from (\ref{rwod_total_power_cons}).

It is easy to check that to achieve the optimal value of problem (\ref{rwod_link_trans_prb})--(\ref{rwod_link_trans_prb_con2}), constraint (\ref{rwod_link_trans_prb_con1}) must be satisfied with equality.
Thus it follows that the optimal transmission probability $p^*=\frac{2{{P}_{\text{C}}}}{{{P}_{\text{S}}}+{{P}_{\text{R}}}+2{\alpha_\text{C}}}$. Hence, problem (\ref{rwod_link_trans_prb})--(\ref{rwod_link_trans_prb_con2}) can be simplified as
\begin{align}
\label{rwod_trans_prb_quasi}
\mathcal{C}_\text{C}\left( {{P}_{\text{C}}} \right)=\underset{\left\{{{P}_{\text{S}}},{{P}_{\text{R}}}\right\}}{\mathop{\max }}\,&\frac{\min \left\{ \mathcal{C}\left( {{P}_{\text{S}}}{{h}_{\text{SR}}} \right),\mathcal{C}\left( {{P}_{\text{R}}}{{h}_{\text{RD}}} \right) \right\}}{{{P}_{\text{S}}}+{{P}_{\text{R}}}+2{\alpha_\text{C}}}\cdot {{P}_{\text{C}}}\\
\label{rwod_trans_prb_quasi_cons1}
\text{s.t. }\ &{{P}_{\text{S}}}+{{P}_{\text{R}}}\ge 2{{P}_{\text{C}}}-2{\alpha_\text{C}},\\
\label{rwod_trans_prb_quasi_cons2}
&{{P}_{\text{S}}}\ge0,\ {{P}_{\text{R}}}\ge0,
\end{align}
where (\ref{rwod_trans_prb_quasi_cons1}) is obtained by substituting $p^*$ into the constraint $0\le p \le 1$.

Since objective function (\ref{rwod_trans_prb_quasi}) is a concave function divided by a linear function, it is quasiconcave over $P_\text{S}$ and $P_\text{R}$.
The maximum value is achieved when $\mathcal{C}\left( {{P}_{\text{S}}}{{h}_{\text{SR}}} \right)=\mathcal{C}\left( {{P}_{\text{R}}}{{h}_{\text{RD}}} \right)$ due to the characteristics of quasiconcave functions.
Thus, substituting ${{P}_{\text{R}}}=\frac{{{P}_{\text{S}}}{{h}_{\text{SR}}}}{{h}_{\text{RD}}}$ into (\ref{rwod_trans_prb_quasi}) and (\ref{rwod_trans_prb_quasi_cons1}), problem (\ref{rwod_trans_prb_quasi})--(\ref{rwod_trans_prb_quasi_cons2}) can be rewritten as
\begin{align}
\label{rwod_trans_prb_quasi_no_min}
\mathcal{C}_\text{C}\left( {{P}_{\text{C}}} \right)=\underset{{{P}_{\text{S}}}\ge0}{\mathop{\max }}\,&\frac{{{h}_{\text{RD}}}\mathcal{C}\left( {{P}_{\text{S}}}{{h}_{\text{SR}}} \right)}{\left( {{h}_{\text{SR}}}+{{h}_{\text{RD}}} \right){{P}_{\text{S}}}+2{{h}_{\text{RD}}}{\alpha_\text{C}}}\cdot {{P}_{\text{C}}}\\
\label{rwod_trans_prb_quasi_no_min_con1}
\text{s.t. }\ &\left( {{h}_{\text{SR}}}+{{h}_{\text{RD}}} \right){{P}_{\text{S}}}\ge 2{{h}_{\text{RD}}}\left( {{P}_{\text{C}}}-{\alpha_\text{C}} \right).
\end{align}

Define
\begin{equation}
{{P}_{\text{ee5}}}\triangleq\mathop{\arg}\underset{{{P}_{\text{S}}}\ge0}{\mathop{\max }}\,\frac{{{P}_{\text{C}}}{{h}_{\text{RD}}}\mathcal{C}\left( {{P}_{\text{S}}}{{h}_{\text{SR}}} \right)}{\left( {{h}_{\text{SR}}}+{{h}_{\text{RD}}} \right){{P}_{\text{S}}}+2{{h}_{\text{RD}}}{\alpha_\text{C}}},
\end{equation}
which achieves the maximum value of (\ref{rwod_trans_prb_quasi_no_min}) without considering constraint (\ref{rwod_trans_prb_quasi_no_min_con1}).
Then, the optimal power allocation of problem (\ref{rwod_origin_prb})--(\ref{rwod_origin_prb_cons}) and the average throughput for RAT-WDL are given in the following proposition.
\begin{proposition}
\label{prp4_CRwoDPC_char}
The optimal power allocation for problem (\ref{rwod_origin_prb})--(\ref{rwod_origin_prb_cons}) is given as: Transmit with power value $\left( P_{\text{S}}^{*},P_{\text{R}}^{*} \right)$ over $p^*$ portion of time slots and keep silent for the rest of slots, where
\begin{equation}
P_{\text{S}}^{*}=\max \left( {{P}_{\text{ee5}}},\frac{2{{h}_{\text{RD}}}}{{{h}_{\text{SR}}}+{{h}_{\text{RD}}}}\left( {{P}_{\text{C}}}-{\alpha_\text{C}} \right) \right),
\end{equation}
\begin{equation}
P_{\text{R}}^{*}=\frac{{{h}_{\text{SR}}}}{{{h}_{\text{RD}}}}P_{\text{S}}^{*},
\end{equation}
and $p^*=\frac{2{{P}_{\text{C}}}}{{{P}_{\text{S}}^*}+{{P}_{\text{R}}^*}+2{\alpha_\text{C}}}$.
With the optimal power allocation, the average throughput $\mathcal{C}_\text{C}\left( {{P}_{\text{C}}} \right)$ defined in (\ref{rwod_origin_prb}) for RAT-WDL is given as
\begin{equation}
\label{CRwoDPC}
\mathcal{C}_\text{C}\left( {{P}_{\text{C}}} \right)=\begin{cases}
   \frac{{{P}_{\text{C}}}{{h}_{\text{RD}}}\mathcal{C}\left( {{P}_{\text{ee5}}}{{h}_{\text{SR}}} \right)}{\left( {{h}_{\text{SR}}}+{{h}_{\text{RD}}} \right){{P}_{\text{ee5}}}+2{{h}_{\text{RD}}}{\alpha_\text{C}}} & 0\le {{P}_{\text{C}}}\le \frac{{{P}_{\text{S}}}\left({{h}_{\text{SR}}}+{{h}_{\text{RD}}}\right)}{2{{h}_{\text{RD}}}}+{\alpha_\text{C}}  \\
   \frac{\mathcal{C}\left( \frac{2{{h}_{\text{SR}}}{{h}_{\text{RD}}}}{{{h}_{\text{SR}}}+{{h}_{\text{RD}}}}\left( {{P}_{\text{C}}}-{\alpha_\text{C}} \right) \right)}{2} & {{P}_{\text{C}}}> \frac{{{P}_{\text{S}}}\left({{h}_{\text{SR}}}+{{h}_{\text{RD}}}\right)}{2{{h}_{\text{RD}}}}+{\alpha_\text{C}}.  \\
\end{cases}
\end{equation}
\end{proposition}
\begin{IEEEproof}
Please see Appendix \ref{prf_prp4_CRwoDPC_char}.
\end{IEEEproof}

\subsection{Asymptotic Analysis}
\label{subsec_RT_AA}
In this subsection, the asymptotic performance for RAT-WDL is analyzed.
\subsubsection{Low SNR Regime}
As ${{P}_{\text{C}}}\to 0 $, the average throughput for RAT-WDL at the low SNR regime is given in (\ref{CRwoDPC}):
\begin{equation}
\mathcal{C}_\text{C}\left( {{P}_{\text{C}}} \right)=
   \frac{{{h}_{\text{RD}}}\mathcal{C}\left( {{P}_{\text{ee5}}}{{h}_{\text{SR}}} \right)}{\left( {{h}_{\text{SR}}}+{{h}_{\text{RD}}} \right){{P}_{\text{ee5}}}+2{{h}_{\text{RD}}}{\alpha_\text{C}}}\cdot {{P}_{\text{C}}}.
\end{equation}

It is interesting to note that $\mathcal{C}_\text{C}\left( {{P}_{\text{C}}} \right)$ is also a linear function of the average power budget ${{P}_{\text{C}}}$ at the low SNR regime.
The scaling factors $\frac{{{h}_{\text{RD}}}\mathcal{C}\left( {{P}_{\text{ee5}}}{{h}_{\text{SR}}} \right)}{\left( {{h}_{\text{SR}}}+{{h}_{\text{RD}}} \right){{P}_{\text{ee5}}}+2{{h}_{\text{RD}}}{\alpha_\text{C}}}$ is the maximum EE for the case of RAT-WDL.

\subsubsection{High SNR Regime}
Based on the results in (\ref{CRwoDPC}), as ${{P}_{\text{C}}}\to \infty $, the average throughput for RAT-WDL at the high SNR regime is asymptotically given as
\begin{equation}
\mathcal{C}_\text{C}\left( {{P}_{\text{C}}} \right)\approx
   \frac{1}{2} \log_2\left( \frac{2{{h}_{\text{SR}}}{{h}_{\text{RD}}}}{{{h}_{\text{SR}}}+{{h}_{\text{RD}}}}\left( {{P}_{\text{C}}}-{\alpha_\text{C}} \right) \right).
\end{equation}

Note that $\frac{2{{h}_{\text{SR}}}{{h}_{\text{RD}}}}{{{h}_{\text{SR}}}+{{h}_{\text{RD}}}}\left( {{P}_{\text{C}}}-{\alpha_\text{C}} \right)$ is a linear function of $P_\text{C}$.
The reciprocals of $\mathcal{C}_\text{C}\left( {{P}_{\text{C}}} \right)$ and $\mathcal{C}_\text{B}\left( {{P}_{\text{B}}} \right)$ are infinitesimal of the same order when the average power budgets approach infinity.
Furthermore, it is obvious that the multiplexing gain of RAT-WDL is also $\frac{1}{2}$.
Thus, RAT-WDL and RAT-DL are of similar performances at the high SNR regime.

\section{Numerical Results}
\label{sec_sim}
In this section, simulations are performed to compare the performances of the proposed optimal power allocation and various suboptimal schemes.
\begin{itemize}
\item DLT: denotes direct link transmission, discussed in Section \ref{subsec_dlt}, whose power allocation is given in Lemma \ref{lem1_direct_link_opt_sol}.
\item RAT-DL: denotes for relay assisted transmission with direct link, discussed in Section \ref{subsec_rat}, whose optimal power allocation is given in Proposition \ref{prp2_CRPB_char}.
\item MT: denotes mixed transmission, discussed in Section \ref{subsec_mt}, whose optimal power allocation is given in Proposition \ref{prp3_mix_opt_sol}.
\item RAT-WDL: denotes relay assisted transmission without direct link, discussed in Section \ref{subsec_rt}, whose optimal power allocation is given in Proposition \ref{prp4_CRwoDPC_char}.
\item CDLT: denotes continuous direct link transmission, which transmits only with the direct link every time slot.
The power allocation for CDLT is given as
\begin{equation}
{P_{\text{S}}}^*=\begin{cases}
   P_0-\alpha_\text{A} &   P_0>\alpha_\text{A}\\
   0 & \text{otherwise}.  \\
\end{cases}
\end{equation}
\item CRAT-DL: denotes continuous relay assisted transmission with direct link, where the source transmits with the help of the relay every time slot.
    The power allocation for CRAT-DL is given as
    \begin{equation}
\left( P_{\text{S}}^{*},P_{\text{R}}^{*} \right)=\begin{cases}
   \left( V,2{{P}_{\text{B}}}-2\alpha_\text{B}-V\right) &  P_0>\alpha_\text{B}\\
   \left( 0,0 \right) & \text{otherwise},  \\
\end{cases}
\end{equation}
\end{itemize}
where $V$ is given in (\ref{v}).
\subsection{Average Power Budgets vs. Throughputs}
\begin{figure}[!t]
\centering
\subfloat[]{\includegraphics[width=3in]{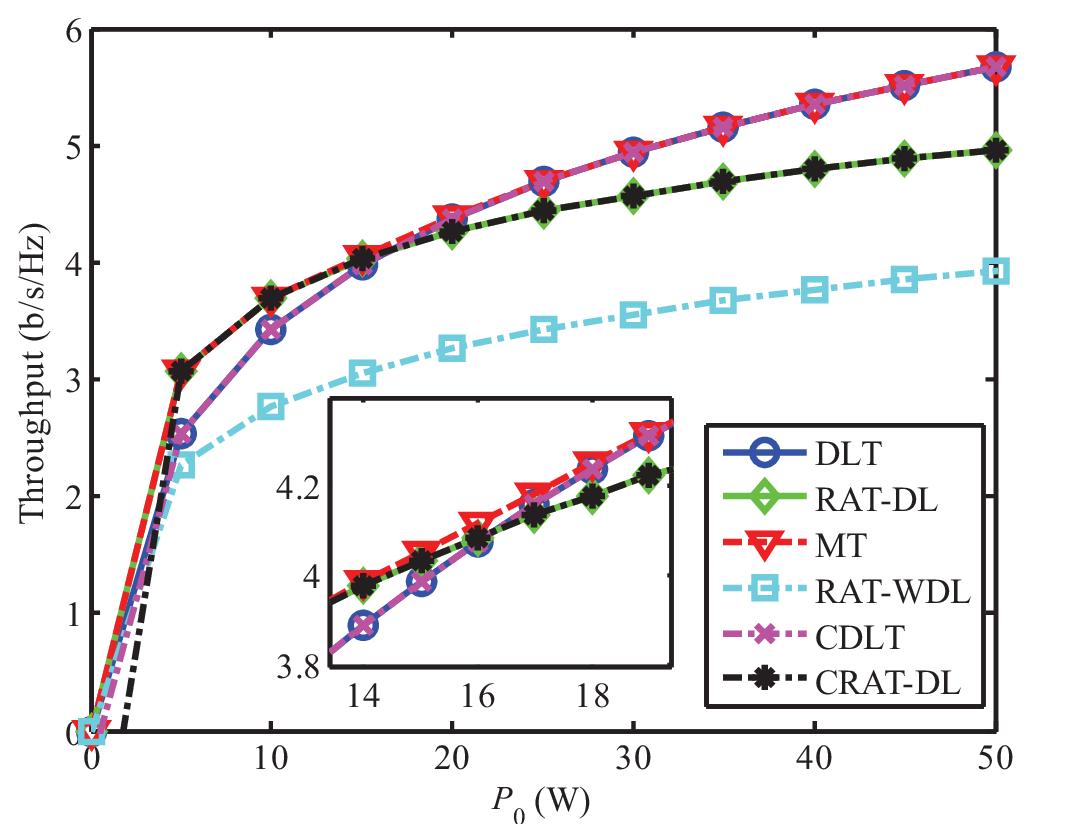}
\label{fig_thphighsnr}}
\hfil
\subfloat[]{\includegraphics[width=3in]{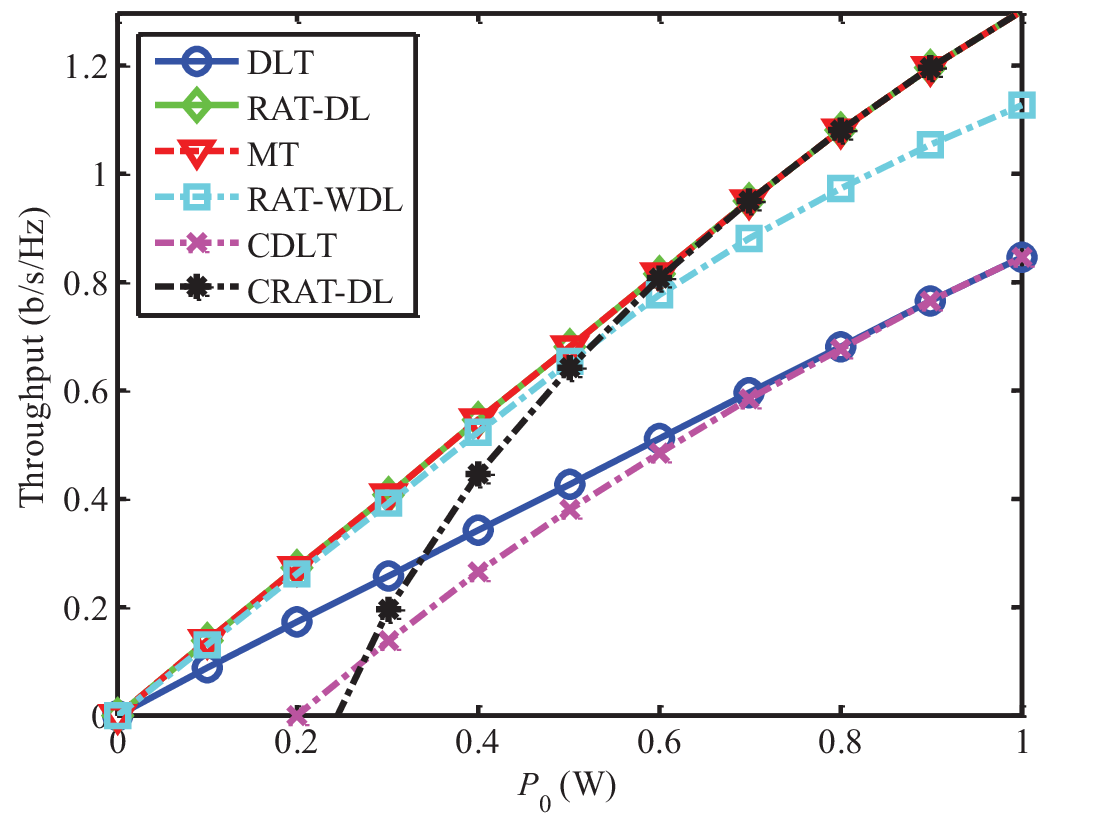}
\label{fig_thplowsnr}}
\caption{Average power budget vs. throughputs in: (a) high SNR regime; (b) low SNR regime.}
\label{fig_transschemescomp}
\end{figure}

Fig. \ref{fig_transschemescomp} compares the performances of several transmission schemes at both high and low SNR regimes.
The circuit power consumptions are set as $\alpha_\text{A}=0.2$ W, $\alpha_\text{B}=0.24$ W, $\alpha_\text{C}=0.18$ W.
The channel gains are set as $h_\text{SD}=1$, $h_\text{SR}=10$, $h_\text{RD}=3$.
It is easy to see that MT always outperforms other transmission schemes.
In Fig. \ref{fig_transschemescomp}(a), when the average power budget $P_0$ is small, throughput curves of MT and RAT-DL coincide; when $P_0$ gets larger, throughput curves of MT, DLT and CDLT coincide.
At the high SNR regime, DLT and CDLT outperform RAT-DL and CRAT-DL, which is due to the multiplexing gain.
Besides, the performance slope of RAT-DL and RAT-WDL are similar, which proves our analysis in Section \ref{subsec_RT_AA}.
Fig. \ref{fig_transschemescomp}(b) depicts the linear parts of $\mathcal{C}_\text{A}\left( {{P}_{\text{A}}} \right)$, $\mathcal{C}_\text{B}\left( {{P}_{\text{B}}} \right)$, and $\mathcal{C}_\text{C}\left( {{P}_{\text{C}}} \right)$ in DLT, RAT-DL, and RAT-WDL.
At the low SNR regime, when $P_0=0.5$ W, throughput performance of RAT-DL/MT is about 0.3 b/s/Hz larger than that of DLT.
Moreover, the performance gap enlarges as $P_0$ increases.
RAT-DL, which coincides with MT, outperforms other transmission schemes.
It suggests that RAT-DL is more energy efficient than DLT and RAT-WDL at the low SNR regime.

\subsection{Channel Gains vs. Mixed Transmission Throughputs}
\begin{figure}[!t]
\centering
\subfloat[]{\includegraphics[width=3in]{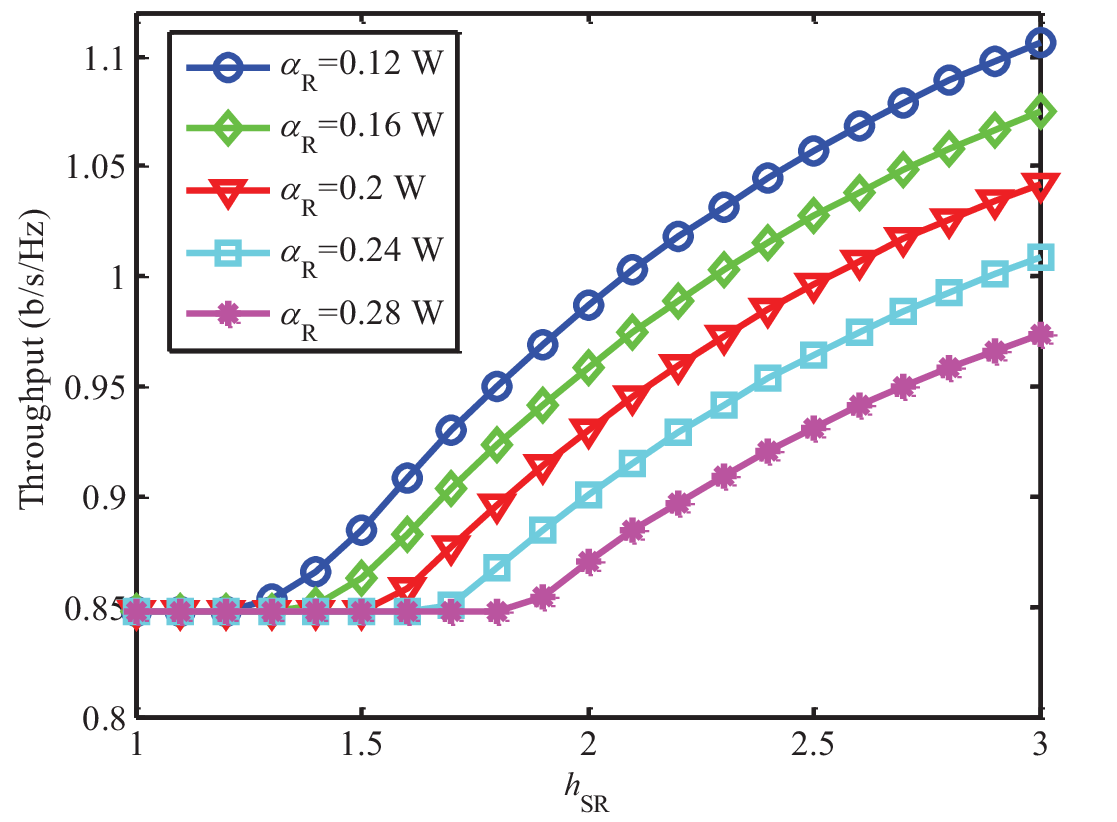}
\label{fig_hsrvsthp}}
\hfil
\subfloat[]{\includegraphics[width=3in]{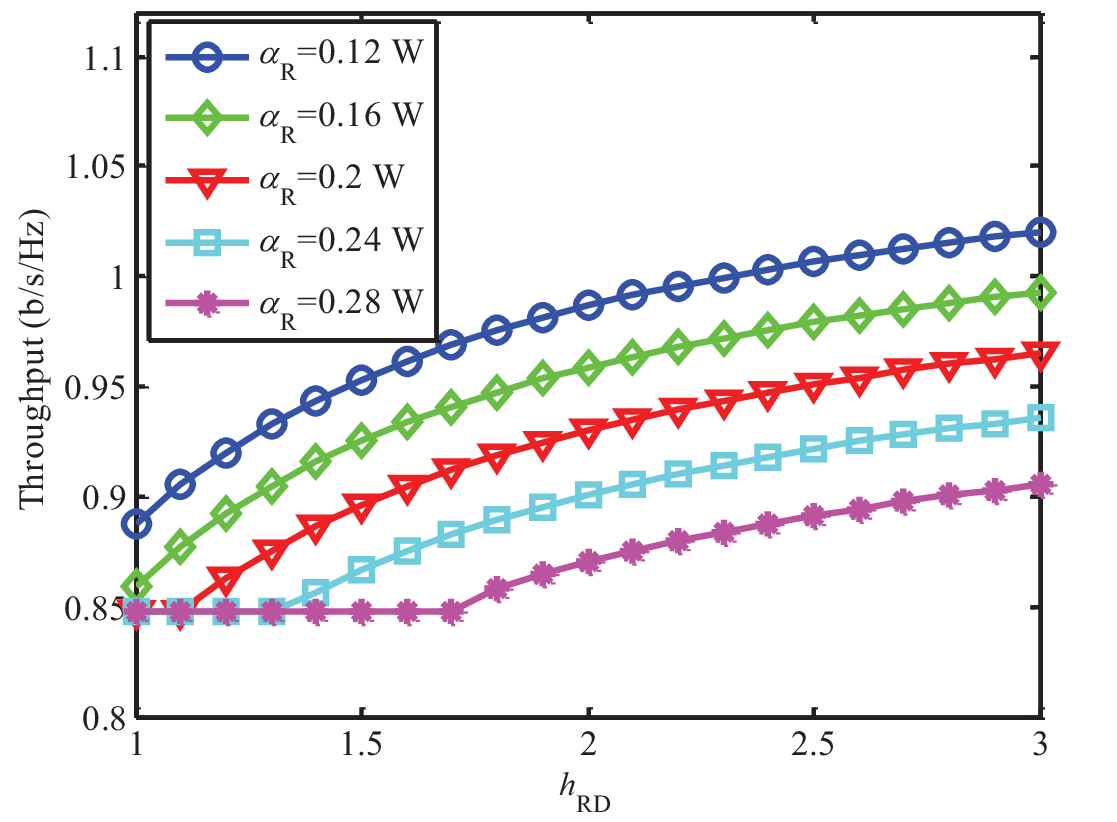}
\label{fig_hrdvsthp}}
\caption{Channel gains vs. mixed transmission throughputs: (a) $h_\text{SR}$ vs. mixed transmission throughput; (b) $h_\text{RD}$ vs. mixed transmission throughput.}
\label{fig_channelvsmt}
\end{figure}
In this subsection, MT throughputs are compared with different channel gains.
The channel gains are set as $h_\text{SD}=1$, $h_\text{RD}=2$ for Fig. \ref{fig_channelvsmt}(a), and $h_\text{SR}=2$ for Fig. \ref{fig_channelvsmt}(b).
The average power budget is set as $P_0=1$ W.
The circuit power consumptions are set as $\alpha_\text{A}=0.2$ W, and $\alpha_\text{B}=0.12,\ 0.16,\ 0.2,\ 0.24,\ 0.28$ W, respectively.

Fig. \ref{fig_channelvsmt} shows that the increase in $h_\text{SR}$ and $h_\text{RD}$ leads the optimal transmission type changing from the DLT to firstly MT, and then RAT-DL.
It is due to that the throughput of RAT-DL improves as $h_\text{SR}$ and $h_\text{RD}$ increases.
The MT range is too short to be seen in the figures, which is located near the turning point.
Besides, it can be concluded from the figures that as $\alpha_\text{B}$ increases, larger channel gains $h_\text{SR}$ or $h_\text{RD}$ are required for the optimal transmission scheme to choose RAT-DL.
Furthermore, the unit throughput improvement by $h_\text{SR}$ is larger than $h_\text{RD}$ when the optimal transmission type is RAT-DL.

\subsection{Circuit Power Consumptions vs. Mixed Transmission Throughputs}
\begin{figure}[!t]
\centering
\subfloat[]{\includegraphics[width=3in]{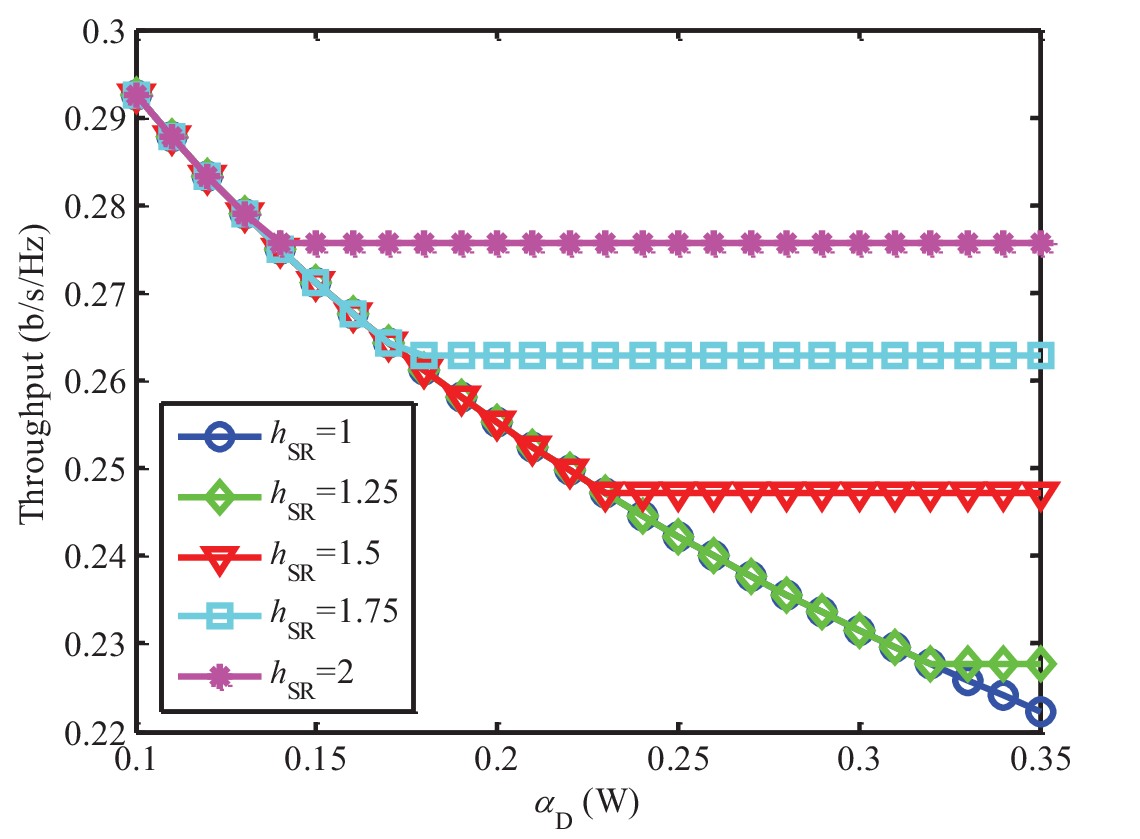}
\label{fig_alphadvsthp}}
\hfil
\subfloat[]{\includegraphics[width=3in]{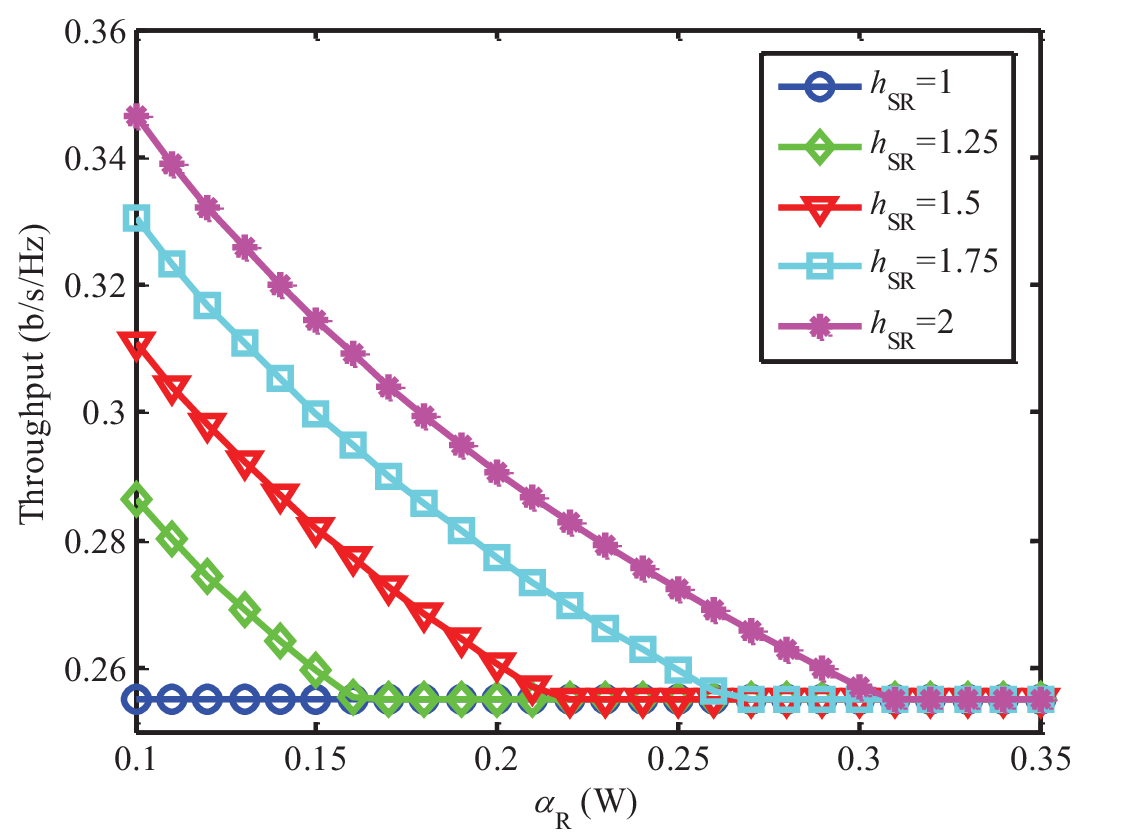}
\label{fig_alpharvsthp}}
\caption{Circuit power consumptions vs. mixed transmission throughputs: (a) $\alpha_\text{A}$ vs. mixed transmission throughput; (b) $\alpha_\text{B}$ vs. mixed transmission throughput.}
\label{fig_circuitvsmt}
\end{figure}
In this subsection, MT throughputs are compared with different circuit power consumptions.
The channel gains are set as $h_\text{SD}=1$, $h_\text{RD}=2$, and $h_\text{SR}=1,\ 1.25,\ 1.5,\ 1.75,\ 2$, respectively.
The average power budget is set as $P_0=0.3$ W.

Fig. \ref{fig_circuitvsmt}(a) shows that the increase in $\alpha_\text{A}$ leads the optimal transmission type changing from the DLT to firstly the MT, and then RAT-DL.
It is due to that the throughput of DLT deteriorates as $\alpha_\text{A}$ increases.
Fig. \ref{fig_circuitvsmt}(b) shows that the increase in $\alpha_\text{B}$ leads the optimal transmission type changing from RAT-DL to firstly MT, and then DLT.
It is due to that the throughput of RAT-DL deteriorates as $\alpha_\text{B}$ increases.
The MT range is too short to be seen in the figures, which is located near the turning point.
Besides, it can be concluded from the figures that as $h_\text{SR}$ increases, the optimal transmission type will change to RAT-DL with fewer circuit power consumption $\alpha_\text{A}$ for Fig. \ref{fig_circuitvsmt}(a), and the optimal transmission type will change to DLT with larger circuit power consumption $\alpha_\text{B}$ for Fig. \ref{fig_circuitvsmt}(b).

\subsection{Optimal Transmission Regions}
\begin{figure}[!t]
\centering
\subfloat[]{\includegraphics[width=3in]{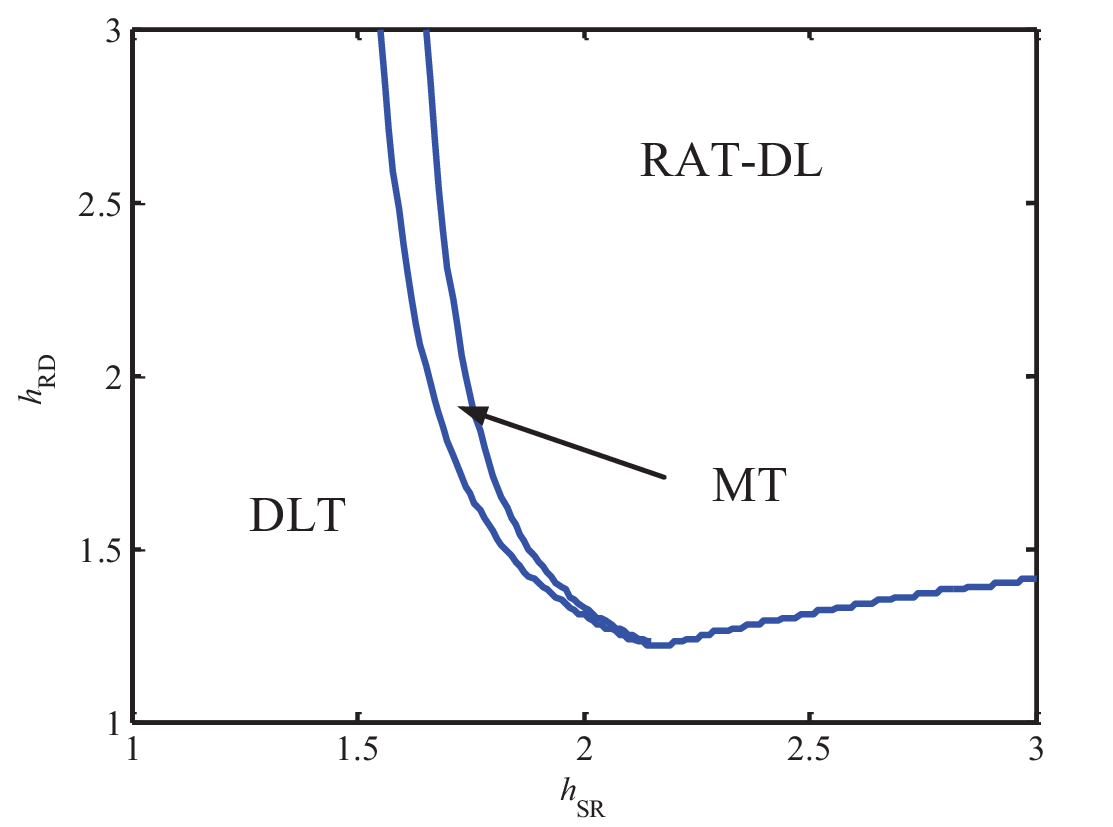}
\label{fig_hsrvshrdp1}}
\hfil
\subfloat[]{\includegraphics[width=3in]{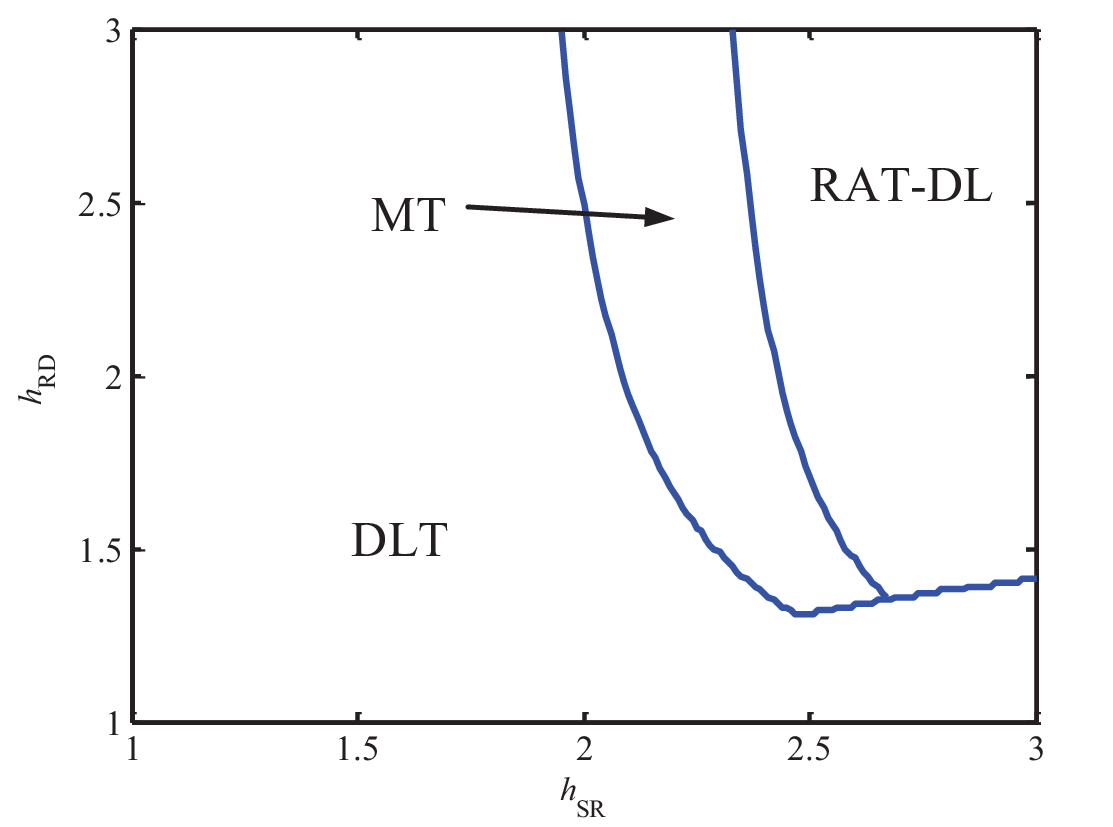}
\label{fig_hsrvshrdp2}}
\caption{Optimal transmission regions with different $P_0$: (a) $P_0=1$ W; (b) $P_0=2$ W.}
\label{fig_optregion}
\end{figure}
In this subsection, the optimal transmission regions are compared with different average power budgets $P_0=1$ W and $P_0=2$ W.
The circuit power consumptions are set as $\alpha_\text{A}=0.2$ W, $\alpha_\text{B}=0.24$ W.
The channel gain for the direct link is set as $h_\text{SD}=1$.

Fig. \ref{fig_optregion} shows that as $P_0$ increases, the regions of DLT and MT expand, while the region of RAT-DL shrinks.
It is due to the multiplexing gain loss of RAT-DL at the high SNR regime, which proves our analysis in Section \ref{subsec_mt_aa}.

\section{Conclusion}
\label{sec_conclu}
In this paper, the throughput optimal power allocation for a basic three-node relay channel with non-ideal circuit power was studied.
Two special scenarios for DLT and RAT-DL were firstly investigated, and their corresponding average throughputs and characteristics were derived.
Then, with the results from these two special cases, the optimal power allocation for the case with direct link was studied, which turns out to be either a single type of transmission (DLT or RAT-DL) or a time sharing of both transmissions according to specific average power budget.
Asymptotic analysis was also given to support the results.
At last, the optimal power allocation for RAT-WDL was analyzed.
Numerical results showed that the proposed optimal power allocation outperforms other suboptimal schemes.

\appendices
\section{Proof of Proposition \ref{prp1_CDPA_char}}
\label{prf_prp1_CDPA_char}
First, the average throughput $\mathcal{C}_\text{A}\left(P_\text{A}\right)$ is obtained by taking $P_{\text{S}}^{*}=\max \left( {{P}_{\text{ee1}}},{{P}_{\text{A}}}-\alpha_\text{A} \right)$ and ${{p}^{*}}=\frac{{{P}_{\text{A}}}}{P_{\text{S}}^{*}+\alpha_\text{A}}$ into the objective function of problem (\ref{d_link_origin_prb}).
Next, it is easy to prove that $\mathcal{C}_\text{A}\left(P_\text{A}\right)$ in (\ref{CDPA}) is continuous over $P_\text{A}\ge 0$ by definition in terms of limits of functions.

Then, examine the differentiability of $\mathcal{C}_\text{A}\left(P_\text{A}\right)$.
Since $\mathcal{C}_\text{A}\left(P_\text{A}\right)$ in (\ref{CDPA}) is obviously differentiable except for the breakpoint $P_\text{A}=P_\text{ee1}+\alpha_\text{A}$, only the differentiability at the breakpoint $P_\text{A}={{P}_{\text{ee1}}}+\alpha_\text{A}$ needs to be discussed.
It is easy to check that
\begin{align}
\nonumber
&\underset{a\to {{0}^{-}}}{\mathop{\lim }}\,\frac{\mathcal{C}_\text{A}\left( {{P}_{\text{ee1}}}+\alpha_\text{A}+a \right)-\mathcal{C}_\text{A}\left( {{P}_{\text{ee1}}}+\alpha_\text{A} \right)}{a}\\
\label{apdx_cdpa_diff1}
=&\underset{a\to {{0}^{-}}}{\mathop{\lim }}\,\mathcal{C}_{_{\text{A}}}^{'}\left( {{P}_{\text{ee1}}}+\alpha_\text{A}+a \right)=\frac{\mathcal{C}\left( {{P}_{\text{ee1}}}{{h}_{\text{SD}}} \right)}{{{P}_{\text{ee1}}}+\alpha_\text{A}},\\
\nonumber
&\underset{a\to {{0}^{+}}}{\mathop{\lim }}\,\frac{\mathcal{C}_\text{A}\left( {{P}_{\text{ee1}}}+\alpha_\text{A}+a \right)-\mathcal{C}_\text{A}\left( {{P}_{\text{ee1}}}+\alpha_\text{A} \right)}{a}\\
\label{apdx_cdpa_diff2}
=&\underset{a\to {{0}^{+}}}{\mathop{\lim }}\,\mathcal{C}_{_{\text{A}}}^{'}\left( {{P}_{\text{ee1}}}+\alpha_\text{A}+a \right)=\frac{{{h}_{\text{SD}}}}{\ln 2\left( 1+{{P}_{\text{ee1}}}{{h}_{\text{SD}}} \right)}.
\end{align}
According to (\ref{apdx_cdpa_diff1}) and (\ref{apdx_cdpa_diff2}), $\frac{\mathcal{C}\left( {{P}_{\text{ee1}}}{{h}_{\text{SD}}} \right)}{{{P}_{\text{ee1}}}+\alpha_\text{A}}=\frac{{{h}_{\text{SD}}}}{\ln 2\left( 1+{{P}_{\text{ee1}}}{{h}_{\text{SD}}} \right)}$ has to hold for differentiability at the breakpoint $P_\text{A}={{P}_{\text{ee1}}}+\alpha_\text{A}$.
Since ${{P}_{\text{ee1}}}\triangleq\underset{{{P}_{\text{S}}}>0}{\mathop{\max }}\,\frac{\mathcal{C}\left( {{P}_{\text{S}}}{{h}_{\text{SD}}} \right)}{{{P}_{\text{S}}}+\alpha_\text{A}}$, it is easy to obtain
\begin{equation}
\frac{d}{dP_\text{S}}{{\left. {{\left[ \frac{\mathcal{C}\left( {{P}_{\text{S}}}{{h}_{\text{SD}}} \right)}{{{P}_{\text{S}}}+\alpha_\text{A}} \right]}} \right|}_{{{P}_{\text{S}}}={{P}_{\text{ee1}}}}}=0,
\end{equation}
which is equivalent to
\begin{equation}
\label{apdx_fermat}
\frac{{{h}_{\text{SD}}}}{\ln 2\left( 1+{{P}_{\text{ee1}}}{{h}_{\text{SD}}} \right)}=\frac{\mathcal{C}\left( {{P}_{\text{ee1}}}{{h}_{\text{SD}}} \right)}{{{P}_{\text{ee1}}}+\alpha_\text{A}}.
\end{equation}
Substituting (\ref{apdx_fermat}) into (\ref{apdx_cdpa_diff2}) leads to
\begin{align}
\nonumber
&\underset{a\to {{0}^{-}}}{\mathop{\lim }}\,\frac{\mathcal{C}_\text{A}\left( {{P}_{\text{ee1}}}+\alpha_\text{A}+a \right)-\mathcal{C}_\text{A}\left( {{P}_{\text{ee1}}}+\alpha_\text{A} \right)}{a}\\
=&\underset{a\to {{0}^{+}}}{\mathop{\lim }}\,\frac{\mathcal{C}_\text{A}\left( {{P}_{\text{ee1}}}+\alpha_\text{A}+a \right)-\mathcal{C}_\text{A}\left( {{P}_{\text{ee1}}}+\alpha_\text{A} \right)}{a},
\end{align}
and it follows that $\mathcal{C}_\text{A}\left(P_\text{A}\right)$ is differentiable at the breakpoint $P_\text{A}={{P}_{\text{ee1}}}+\alpha_\text{A}$.
Thus, $\mathcal{C}_\text{A}\left(P_\text{A}\right)$ is differentiable when $P_\text{A}\ge0$.

At last, examine the concavity of $\mathcal{C}_\text{A}\left(P_\text{A}\right)$.
Since $\mathcal{C}_\text{A}\left(P_\text{A}\right)$ is continuous and differentiable over $P_\text{A}\ge0$, and first-order condition is satisfied by the definition of $\mathcal{C}_\text{A}\left(P_\text{A}\right)$ in (\ref{CDPA}), it is a concave function \cite{SBoyd04}.

Based on the above analysis, Proposition \ref{prp1_CDPA_char} is proved.

\section{Proof of Proposition \ref{prp2_CRPB_char}}
\label{prf_prp2_CRPB_char}
First, the average throughput $\mathcal{C}_\text{B}\left(P_\text{B}\right)$ is obtained.
By taking the optimal solutions given in (\ref{optimal_sol_CRPB}) into (\ref{r_link_trans_prb_quasi_no_min}), the average throughput $\mathcal{C}_\text{B}\left(P_\text{B}\right)$ can be obtained as shown in Proposition \ref{prp2_CRPB_char}.

Next, the continuity and differentiability of $\mathcal{C}_\text{B}\left(P_\text{B}\right)$ is examined.
Note that $\mathcal{C}_\text{B}\left(P_\text{B}\right)$ in (\ref{crpb}) is continuous and differentiable except for the breakpoints
\begin{align}
\nonumber
P_\text{B}=&\left\{\frac{{{P}_{\text{ee2}}}+{{P}_{\text{ee3}}}}{2}+\alpha_\text{B},\ \frac{P_\text{ee4}\left( {{h}_{\text{SR}}}-{{h}_{\text{SD}}} \right)}{{{h}_{\text{RD}}}\left( 1+P_\text{ee4}{{h}_{\text{SD}}} \right)}+\frac{P_\text{ee4}}{2}+\alpha_\text{B},\right.\\
&\left.\frac{h_\text{SD}V^2}{2}+\frac{UV}{h_\text{RD}}+\alpha_\text{B}\right\},
\end{align}
where $U$ and $V$ are given in (\ref{u}) and (\ref{v}), respectively.

First, examine the continuity at the breakpoint $P_\text{B}=\frac{{{P}_{\text{ee2}}}+{{P}_{\text{ee3}}}}{2}+\alpha_\text{B}$.
It is easy to check that
\begin{equation}
\label{apdx_prp2_ctnt1}
\underset{{{P}_{\text{B}}}\to {{\left(\frac{{{P}_{\text{ee2}}}+{{P}_{\text{ee3}}}}{2}+\alpha_\text{B}\right)}^{-}}}{\mathop{\lim }}\,\mathcal{C}_\text{B}\left( {{P}_{\text{B}}} \right)=\frac{ \mathcal{C}\left( {{P}_{\text{ee2}}}{{h}_{\text{SD}}} \right)+\mathcal{C}\left( {{P}_{\text{ee3}}}{{h}_{\text{RD}}} \right) }{2},
\end{equation}
\begin{align}
\label{apdx_prp2_ctnt2}
\nonumber
&\underset{{{P}_{\text{B}}}\to {{\left(\frac{{{P}_{\text{ee2}}}+{{P}_{\text{ee3}}}}{2}+\alpha_\text{B}\right)}^{+}}}{\mathop{\lim }}\,\mathcal{C}_\text{B}\left( {{P}_{\text{B}}} \right)\\
=&\frac{1}{2}\mathcal{C}\left( \frac{{{h}_{\text{SD}}}}{2{{h}_{\text{RD}}}}+\left( \frac{{{P}_{\text{ee2}}}+{{P}_{\text{ee3}}}}{2} \right){{h}_{\text{SD}}}-\frac{1}{2} \right)\\
+&\frac{1}{2}\mathcal{C}\left( \frac{{{h}_{\text{RD}}}}{2{{h}_{\text{SD}}}}+\left( \frac{{{P}_{\text{ee2}}}+{{P}_{\text{ee3}}}}{2} \right){{h}_{\text{RD}}}-\frac{1}{2} \right).
\end{align}

To check the continuity of $\mathcal{C}_\text{B}\left(P_\text{B}\right)$ at the breakpoints $P_\text{B}=\frac{{{P}_{\text{ee2}}}+{{P}_{\text{ee3}}}}{2}+\alpha_\text{B}$, the right-hand sides of (\ref{apdx_prp2_ctnt1}) and (\ref{apdx_prp2_ctnt2}) must be equal.
Denote $f\left( {{P}_{\text{S}}},{{P}_{\text{R}}} \right)=\frac{\mathcal{C}\left( {{P}_{\text{S}}}{{h}_{\text{SD}}} \right)+\mathcal{C}\left( {{P}_{\text{R}}}{{h}_{\text{RD}}} \right)}{{{P}_{\text{S}}}+{{P}_{\text{R}}}+2\alpha_\text{B}}$ for purpose of exposition.
Since $\left( {{P}_{\text{ee2}}},{{P}_{\text{ee3}}} \right)=\underset{\left\{{{P}_{\text{S}}}>0,{{P}_{\text{R}}}>0\right\}}{\mathop{\max }}\,f\left( {{P}_{\text{S}}},{{P}_{\text{R}}} \right)$, it is easy to obtain that ${{f}^{'}_{{{P}_{\text{S}}}}}\left( {{P}_{\text{ee2}}},{{P}_{\text{ee3}}} \right)=0$ and ${{f}^{'}_{{{P}_{\text{R}}}}}\left( {{P}_{\text{ee2}}},{{P}_{\text{ee3}}} \right)=0$, which are equivalent to
\begin{eqnarray}
\label{apdx_prp2_ctnt3}
\frac{{{h}_{\text{SD}}}\left( {{P}_{\text{ee2}}}+{{P}_{\text{ee3}}}+2\alpha_\text{B} \right)}{\ln 2\left( 1+{{P}_{\text{ee2}}}{{h}_{\text{SD}}} \right)}=\mathcal{C}\left( {{P}_{\text{ee2}}}{{h}_{\text{SD}}} \right)+\mathcal{C}\left( {{P}_{\text{ee3}}}{{h}_{\text{RD}}} \right), \\
\label{apdx_prp2_ctnt4}
\frac{{{h}_{\text{RD}}}\left( {{P}_{\text{ee2}}}+{{P}_{\text{ee3}}}+2\alpha_\text{B} \right)}{\ln 2\left( 1+{{P}_{\text{ee3}}}{{h}_{\text{RD}}} \right)}=\mathcal{C}\left( {{P}_{\text{ee2}}}{{h}_{\text{SD}}} \right)+\mathcal{C}\left( {{P}_{\text{ee3}}}{{h}_{\text{RD}}} \right).
\end{eqnarray}
Substituting (\ref{apdx_prp2_ctnt3}) into (\ref{apdx_prp2_ctnt4}) leads to
\begin{equation}
\label{apdx_prp2_ctnt5}
\frac{{{h}_{\text{RD}}}}{ 1+{{P}_{\text{ee3}}}{{h}_{\text{RD}}}}=\frac{{{h}_{\text{SD}}}}{ 1+{{P}_{\text{ee2}}}{{h}_{\text{SD}}}},
\end{equation}
which is equivalent to
\begin{equation}
\label{apdx_prp2_ctnt6}
{{h}_{\text{RD}}}+{{P}_{\text{ee2}}}{{h}_{\text{SD}}}{{h}_{\text{RD}}}={{h}_{\text{SD}}}+{{P}_{\text{ee3}}}{{h}_{\text{SD}}}{{h}_{\text{RD}}}.
\end{equation}
It is easy to obtain the following two equations from (\ref{apdx_prp2_ctnt6}):
\begin{eqnarray}
\label{apdx_prp2_ctnt7}
\frac{{{h}_{\text{SD}}}}{2{{h}_{\text{RD}}}}+\left( \frac{{{P}_{\text{ee2}}}+{{P}_{\text{ee3}}}}{2} \right){{h}_{\text{SD}}}-\frac{1}{2}={{P}_{\text{ee2}}}{{h}_{\text{SD}}},\\
\label{apdx_prp2_ctnt8}
\frac{{{h}_{\text{RD}}}}{2{{h}_{\text{SD}}}}+\left( \frac{{{P}_{\text{ee2}}}+{{P}_{\text{ee3}}}}{2} \right){{h}_{\text{RD}}}-\frac{1}{2}={{P}_{\text{ee3}}}{{h}_{\text{RD}}}.
\end{eqnarray}
Substituting (\ref{apdx_prp2_ctnt7})(\ref{apdx_prp2_ctnt8}) into (\ref{apdx_prp2_ctnt2}) leads to
\begin{align}
\nonumber
&\underset{{{P}_{\text{B}}}\to {{\left(\frac{{{P}_{\text{ee2}}}+{{P}_{\text{ee3}}}}{2}+\alpha_\text{B}\right)}^{-}}}{\mathop{\lim }}\,\mathcal{C}_\text{B}\left( {{P}_{\text{B}}} \right)=\underset{{{P}_{\text{B}}}\to {{\left(\frac{{{P}_{\text{ee2}}}+{{P}_{\text{ee3}}}}{2}+\alpha_\text{B}\right)} ^{+}}}{\mathop{\lim }}\,\mathcal{C}_\text{B}\left( {{P}_{\text{B}}} \right)\\
=&\mathcal{C}_\text{B}\left( \frac{{{P}_{\text{ee2}}}+{{P}_{\text{ee3}}}}{2}+\alpha_\text{B} \right),
\end{align}
and it follows that $\mathcal{C}_\text{B}\left(P_\text{B}\right)$ is continuous at the breakpoint $P_\text{B}=\frac{{{P}_{\text{ee2}}}+{{P}_{\text{ee3}}}}{2}+\alpha_\text{B}$.

Next examine the differentiability of $\mathcal{C}_\text{B}\left(P_\text{B}\right)$ at the breakpoint $P_\text{B}=\frac{{{P}_{\text{ee2}}}+{{P}_{\text{ee3}}}}{2}+\alpha_\text{B}$.
It is easy to check that
\begin{align}
\label{apdx_prp2_dfftb1}
\nonumber
&\underset{a\to {{0}^{-}}}{\mathop{\lim }}\,\frac{\mathcal{C}_\text{B}\left( \frac{{{P}_{\text{ee2}}}+{{P}_{\text{ee3}}}}{2}+\alpha_\text{B}+a\right)-\mathcal{C}_\text{B}\left( \frac{{{P}_{\text{ee2}}}+{{P}_{\text{ee3}}}}{2}+\alpha_\text{B} \right)}{a}\\
\nonumber
=&\underset{a\to {{0}^{-}}}{\mathop{\lim }}\,\mathcal{C}_{_{\text{R}}}^{'}\left( \frac{{{P}_{\text{ee2}}}+{{P}_{\text{ee3}}}}{2}+\alpha_\text{B}+a \right)\\
=&\frac{\mathcal{C}\left( {{P}_{\text{ee2}}}{{h}_{\text{SD}}} \right)+\mathcal{C}\left( {{P}_{\text{ee3}}}{{h}_{\text{RD}}} \right)}{{{P}_{\text{ee2}}}+{{P}_{\text{ee3}}}+2\alpha_\text{B}},
\end{align}
\begin{align}
\label{apdx_prp2_dfftb2}
\nonumber
&\underset{a\to {{0}^{+}}}{\mathop{\lim }}\,\frac{\mathcal{C}_\text{B}\left( \frac{{{P}_{\text{ee2}}}+{{P}_{\text{ee3}}}}{2}+\alpha_\text{B}+a\right)-\mathcal{C}_\text{B}\left( \frac{{{P}_{\text{ee2}}}+{{P}_{\text{ee3}}}}{2}+\alpha_\text{B} \right)}{a}\\
\nonumber
=&\underset{a\to {{0}^{+}}}{\mathop{\lim }}\,\mathcal{C}_{_{\text{R}}}^{'}\left( \frac{{{P}_{\text{ee2}}}+{{P}_{\text{ee3}}}}{2}+\alpha_\text{B}+a \right)\\
=&\frac{1}{2\ln 2}\left( \frac{{{h}_{\text{SD}}}}{\frac{1}{2}+\frac{{{h}_{\text{SD}}}}{2{{h}_{\text{RD}}}}+\left( \frac{{{P}_{\text{ee2}}}+{{P}_{\text{ee3}}}}{2} \right){{h}_{\text{SD}}}}\right.\\
\nonumber
&+\left.\frac{{{h}_{\text{RD}}}}{\frac{1}{2}+\frac{{{h}_{\text{RD}}}}{2{{h}_{\text{SD}}}}+\left( \frac{{{P}_{\text{ee2}}}+{{P}_{\text{ee3}}}}{2} \right){{h}_{\text{RD}}}} \right).
\end{align}
Substituting (\ref{apdx_prp2_ctnt3})(\ref{apdx_prp2_ctnt4})(\ref{apdx_prp2_ctnt7})(\ref{apdx_prp2_ctnt8}) into (\ref{apdx_prp2_dfftb2}) leads to
\begin{align}
\nonumber
&\underset{a\to {{0}^{-}}}{\mathop{\lim }}\,\frac{\mathcal{C}_\text{B}\left( \frac{{{P}_{\text{ee2}}}+{{P}_{\text{ee3}}}}{2}+\alpha_\text{B}+a\right)-\mathcal{C}_\text{B}\left( \frac{{{P}_{\text{ee2}}}+{{P}_{\text{ee3}}}}{2}+\alpha_\text{B} \right)}{a}\\
=&\underset{a\to {{0}^{+}}}{\mathop{\lim }}\,\frac{\mathcal{C}_\text{B}\left( \frac{{{P}_{\text{ee2}}}+{{P}_{\text{ee3}}}}{2}+\alpha_\text{B}+a\right)-\mathcal{C}_\text{B}\left( \frac{{{P}_{\text{ee2}}}+{{P}_{\text{ee3}}}}{2}+\alpha_\text{B} \right)}{a},
\end{align}
and it follows that $\mathcal{C}_\text{B}\left(P_\text{B}\right)$ is differentiable at the breakpoint $P_\text{B}=\frac{{{P}_{\text{ee2}}}+{{P}_{\text{ee3}}}}{2}+\alpha_\text{B}$.

The continuity and differentiability at the other two breakpoints can be examined in similar ways, the proof is omitted due to space limitations.
Based on the above analysis, $\mathcal{C}_\text{B}\left(P_\text{B}\right)$ is continuous and differentiable when $P_\text{B}\ge0$.
Besides, since first-order condition is satisfied by the definition of $\mathcal{C}_\text{B}\left(P_\text{B}\right)$, it is a concave function \cite{SBoyd04}.

Thus, Proposition \ref{prp2_CRPB_char} is proved.

\section{Proof of Proposition \ref{prp3_mix_opt_sol}}
\label{prf_prp3_mix_opt_sol}
It is worth to note that the objective function (\ref{m_link_origin_prb}) is $\theta \mathcal{C}_\text{A}\left( {{P}_{\text{A}}} \right)+\left( 1-\theta  \right)\mathcal{C}_\text{B}\left( {{P}_{\text{B}}} \right)$, which stands for any line segments between any two points on line $\mathcal{C}_\text{A}\left(P\right)$ and $\mathcal{C}_\text{B}\left(P\right)$.
Constraint (\ref{m_link_origin_prb_con1}) gives the relationship between $P_0$ and $\left\{ P_\text{A},\ P_\text{B},\ \theta \right\}$.
Thus, the optimization problem can be interpreted as finding the maximum value of any line segments between any two points on line $\mathcal{C}_\text{A}\left(P\right)$ and $\mathcal{C}_\text{B}\left(P\right)$ with specific x-coordinate $P_0$.

In Case 1 and Case 2, $\mathcal{C}_\text{A}\left(P\right) \ge \mathcal{C}_\text{B}\left(P\right)$ over $P>0$, i.e., the line segments are upper bounded by $\mathcal{C}_\text{A}\left(P\right)$.
Thus, it is obvious in these cases that $\left( P_{\text{A}}^{*},P_{\text{B}}^{*} \right)=\left( P_0,0 \right)$.
Besides, $\theta^{*}$ can be obtained by substituting $\left( P_{\text{A}}^{*},P_{\text{B}}^{*} \right)$ into (\ref{m_link_origin_prb_con1}).

In Case 3, as illustrated in Fig. \ref{fig_mixcase}, the domain is divided into several intervals by the x-coordinates of the tangent points.
When $P_0$ falls into the interval where the line segments are upper bounded by $\mathcal{C}_\text{A}\left(P\right)$, the power should all be allocated to DLT;
When $P_0$ falls into the interval where the line segments are upper bounded by $\mathcal{C}_\text{B}\left(P\right)$, the power should all be allocated to RAT;
When $P_0$ falls into the interval where the line segments are upper bounded by the tangent line of $\mathcal{C}_\text{A}\left(P\right)$ and $\mathcal{C}_\text{B}\left(P\right)$, the power value allocated for DLT and RAT are the corresponding x-coordinates of the tangent points on $\mathcal{C}_\text{A}\left(P\right)$ and $\mathcal{C}_\text{B}\left(P\right)$, respectively.

With the above results, the optimal power allocation can be easily obtained according to $P_0$ as shown in Proposition \ref{prp3_mix_opt_sol}, and thus, Proposition \ref{prp3_mix_opt_sol} is proved.

\section{Proof of Proposition \ref{prp4_CRwoDPC_char}}
\label{prf_prp4_CRwoDPC_char}
It is easy to check that objective function (\ref{rwod_trans_prb_quasi_no_min}) is quasiconcave over $P_\text{S}$, since it is a concave function divided by a linear function \cite{SBoyd04}.
It is increasing if $0\le {{P}_{\text{C}}}\le \frac{{{h}_{\text{SR}}}+{{h}_{\text{RD}}}}{2{{h}_{\text{RD}}}}{{P}_{\text{S}}}+{\alpha_\text{C}}$ and decreasing if ${{P}_{\text{C}}}> \frac{{{h}_{\text{SR}}}+{{h}_{\text{RD}}}}{2{{h}_{\text{RD}}}}{{P}_{\text{S}}}+{\alpha_\text{C}}$.
Thus, $P_{\text{S}}^{*}=\max \left( {{P}_{\text{ee5}}},\frac{2{{h}_{\text{RD}}}}{{{h}_{\text{SR}}}+{{h}_{\text{RD}}}}\left( {{P}_{\text{C}}}-{\alpha_\text{C}} \right) \right)$ achieves the maximum value of problem (\ref{rwod_trans_prb_quasi_no_min})--(\ref{rwod_trans_prb_quasi_no_min_con1}).
$P_{\text{R}}^{*}=\frac{{{h}_{\text{SR}}}}{{{h}_{\text{RD}}}}P_{\text{S}}^{*}$ is obtained from ${{P}_{\text{S}}}{{h}_{\text{SR}}}={{P}_{\text{R}}}{{h}_{\text{RD}}}$.
Substituting $P_{\text{S}}^{*}$ and $P_{\text{R}}^{*}$ into objective function (\ref{rwod_trans_prb_quasi_no_min}), the average throughput $\mathcal{C}_\text{C}\left( {{P}_{\text{C}}} \right)$ for RAT-WDL is obtained as (\ref{CRwoDPC}).
Thus, Proposition \ref{prp4_CRwoDPC_char} is proved.

\bibliographystyle{IEEEtran}
\bibliography{IEEEabrv,ref}

\end{document}